\documentclass[
aps,
prb,
reprint,
longbibliography,
superscriptaddress,
nobalancelastpage
]{revtex4-2}

\usepackage[caption=false]{subfig}
\usepackage[T1]{fontenc}
\usepackage[utf8]{inputenc}
\usepackage[english]{babel}

\usepackage{graphicx}
\usepackage{xcolor}
\usepackage{placeins}
\usepackage{adjustbox}

\usepackage{amsmath}
\usepackage{amssymb}
\usepackage{amsfonts}
\usepackage{amsthm}
\usepackage{mathtools}

\usepackage{physics}
\usepackage{booktabs}
\usepackage{multirow}

\usepackage{siunitx}
\usepackage{csquotes}

\usepackage[
    colorlinks=true,
    linkcolor=blue,
    citecolor=blue,
    urlcolor=blue
]{hyperref}

\usepackage[noabbrev,nameinlink]{cleveref}

\begin{document}
\title{Reduction of intrinsic losses in nanomechanical silicon nitride resonators through thermal treatment in ultrahigh vacuum}

\author{Nicola Cavalleri}
\affiliation{Institute of Sensor and Actuator Systems, TU Wien, Gusshausstrasse 27-29, 1040 Vienna, Austria.}
\author{Ariane Giesriegl}
\affiliation{Institute of Sensor and Actuator Systems, TU Wien, Gusshausstrasse 27-29, 1040 Vienna, Austria.}
\author{Robert G. West}
\affiliation{Institute of Sensor and Actuator Systems, TU Wien, Gusshausstrasse 27-29, 1040 Vienna, Austria.}
\author{Kostas Kanellopulos}
\affiliation{Institute of Sensor and Actuator Systems, TU Wien, Gusshausstrasse 27-29, 1040 Vienna, Austria.}
\author{Saeed Rasouli}
\affiliation{Research Unit of Nanoelectronic Devices, TU Wien, Gusshausstrasse 25-25a, 1040 Vienna, Austria.}
\author{Daniele Nazzari}
\affiliation{Research Unit of Nanoelectronic Devices, TU Wien, Gusshausstrasse 25-25a, 1040 Vienna, Austria.}
\author{Pedram Sadeghi}
\affiliation{Institute of Sensor and Actuator Systems, TU Wien, Gusshausstrasse 27-29, 1040 Vienna, Austria.}
\author{Sebastian Alberti}
\affiliation{Institute of Sensor and Actuator Systems, TU Wien, Gusshausstrasse 27-29, 1040 Vienna, Austria.}
\author{Antonius Armanious}
\affiliation{Laboratory for Solid State Physics, ETH Z\"{u}rich, CH-8093 Z\"urich, Switzerland.}
\affiliation{Quantum Center, ETH Zurich, CH-8093 Zurich, Switzerland}
\author{Silvan Schmid}
    \email[Correspondence email address: ]{silvan.schmid@tuwien.ac.at}
    \affiliation{Institute of Sensor and Actuator Systems, TU Wien, Gusshausstrasse 27-29, 1040 Vienna, Austria.}

\date{\today} 

\begin{abstract}
Since the discovery of dissipation dilution, silicon nitride (SiN) nanomechanical resonators have set the benchmark for ultracoherent mechanical systems, with geometry and strain engineering driving remarkable gains in the $f \cdot Q$ product. Surface loss, however, has remained the dominant and largely unaddressed dissipation channel. Here, we demonstrate a geometry-independent approach that directly targets surface loss: thermal treatment in ultrahigh vacuum. Treatment at \SI{1000}{\celsius} enhances the intrinsic quality factor of dissipation-diluted SiN membrane resonators by up to a factor of 20, reduces the surface loss eightfold, and simultaneously increases the tensile stress. Photothermal infrared spectroscopy and \textit{in situ} X-ray photoelectron spectroscopy trace the enhancement to thermally activated silanol condensation --- the conversion of surface hydroxyl terminations into siloxane bridges --- and the reversibility of both quality factor and stress under controlled humidity confirms the surface-chemical origin. These results establish surface chemistry as a tunable parameter for next-generation ultracoherent nanomechanical resonators.
\end{abstract}

\maketitle

\section{Introduction}

Silicon nitride (SiN) thin films have enabled the development of ultrahigh-$Q$ micro- and nanomechanical resonators over the past two decades, beginning with the discovery of dissipation dilution in nanomechanical SiN string resonators by Verbridge et al.\ in 2006~\cite{verbridge2006}. 
Dissipation dilution arises because the process-induced tensile stress effectively adds a nearly lossless restoring force, thereby diluting the influence of intrinsic material damping on the overall quality factor~\cite{saulson1994,schmid2008dampingpolimers,unterreithmeier2010, schmid2011damping, yu2012control}.
This mechanism allows stressed SiN resonators to reach quality factors on the order of~$10^6$ at room temperature, despite their large surface-to-volume ratios~\cite{engelsen2024ultrahigh}.

High mechanical quality factors are central to a wide range of sensing and fundamental-physics applications, including ultrasensitive force detection~\cite{membrane_SFM_ETH_2021}
and cavity optomechanics~\cite{Aspelmeyer_cavity_2014}.
In particular, achieving quantum-coherent operation of a mechanical resonator at bath temperature~$T$ requires the frequency--quality factor product to satisfy $f_0 \cdot Q \gtrsim k_B T / 2\pi\hbar$, corresponding to $f_0 \cdot Q \gtrsim \SI{6e12}{\hertz}$ at room temperature~\cite{quantum_Opto_RT_norte2016}.
Reaching this threshold would eliminate the need for cryogenic cooling and the associated experimental overhead~\cite{beccari2022strained}, making room-temperature quantum optomechanics a realistic prospect~\cite{quantum_Opto_RT_norte2016, huang2024room}.

The discovery of dissipation dilution spurred a series of resonator-engineering advances aimed at maximising~$Q$.
Embedding the resonator in a phononic crystal (PnC) suppresses acoustic radiation losses through a bandgap around the mode frequency of interest and, when the PnC is patterned directly into the membrane, additionally creates soft-clamping conditions that allow the resonant mode to decay evanescently into the crystal lattice, thereby eliminating bending losses at the clamping boundary~\cite{Tsaturyan2017_PnC, ghadimi2018}.
Soft-clamped membranes achieved quality factors on the order of~$10^8$ at room temperature~\cite{Tsaturyan2017_PnC}.
Subsequent work on hierarchical geometries~\cite{bereyhi2022} 
and perimeter modes~\cite{bereyhi_perimeter2022} further enhanced dissipation dilution through elastic strain engineering.
Most recently, the combination of these strategies has yielded room-temperature quality factors approaching $8 \times 10^9$~\cite{cupertino2024}.

Underlying all of these results is a common limiting mechanism: surface loss~\cite{Yasumura2000_surfQ,Silvan2014_evidence_of_surf_loss}.
As the film thickness~$h$ enters the sub-micron regime, the intrinsic quality factor scales linearly with thickness, $Q_\mathrm{int}\approx\beta\cdot h$ with $\beta\approx 60$~nm$^{-1}$~\cite{Silvan2014_evidence_of_surf_loss,benga2026determination}, so that thinner films exhibit proportionally lower $Q_\mathrm{int}$.
This scaling is independent of resonator geometry and identifies surface loss as the dominant dissipation channel in all state-of-the-art SiN resonators reported to date~\cite{Tsaturyan2017_PnC, ghadimi2018, bereyhi2022, bereyhi_perimeter2022}.
Analogous surface dissipation has been observed in crystalline silicon resonators~\cite{Yasumura2000_surfQ, SurfQ_Si_Yang2000}, indicating that surface loss is a barrier common to both material platforms.

For the thinnest films, an additional excess-loss channel has recently been introduced to account for the further reduction of $Q_\mathrm{int}$~\cite{benga2026determination}, giving
\begin{equation}
    \frac{1}{Q_\mathrm{int}(h)} =
    \frac{1}{Q_\mathrm{V}^\infty}
    + \frac{6\delta}{h}\,\frac{1}{Q_\mathrm{S}}
    + \frac{1}{Q_\mathrm{V}^{\mathrm{def}}}\,e^{-h/h_\mathrm{def}},
    \label{eq:Qint_surf_limited}
\end{equation}
where $Q_\mathrm{V}^\infty$ is the bulk loss limit, $Q_\mathrm{S}$ the surface-loss quality factor of lossy layers of thickness $\delta$ on both surfaces, and $Q_\mathrm{V}^{\mathrm{def}}$ an excess volume loss whose contribution decays exponentially over a characteristic thickness $h_\mathrm{def}$.
An analogous excess loss in the thinnest films has been reported for silicon carbide thin-film resonators \cite{Romero2020_excesslossSiC}, suggesting that this channel is not specific to SiN. The microscopic origin of the surface loss itself remains unclear;  a native silicon dioxide layer~\cite{Nik2017_plasma_ox}, surface hydroxyl groups~\cite{giesriegl2026hf,ono_ion_dissipation_2005}, and adsorbed hydrocarbons~\cite{Wang2004,tao_permanent_reduction_dissipation_2015} have been suggested as contributing factors.
Thermal treatment at temperatures up to \SI{1000}{\celsius} increases $Q$ by more than an order of magnitude in crystalline silicon and silicon nitride cantilevers~\cite{SurfQ_Si_Yang2000,Yang2002,Yasumura2000_surfQ}.
To our knowledge, no systematic thermal-treatment study of dissipation-diluted SiN resonators has been reported.

In this work, we investigate the nature and mitigation of surface losses in stressed SiN thin-film resonators. 
We characterize the surface composition using nanomechanical photothermal Fourier-transform infrared spectroscopy (FTIR) and X-ray photoelectron spectroscopy (XPS).
We systematically investigate the effects of ultrahigh vacuum (UHV) and thermal treatment on reducing dissipation and the associated changes in film stress. We further assess the persistence of the heating-induced enhancement of the $Q$ upon re-exposure to dry and humid air at room temperature, and compare the results with those obtained for samples stored under high vacuum.

\section{Methods}\label{sec:methods}
 
\subsection{SiN membranes}\label{subsec:samples}
In this work, silicon-rich and stoichiometric low-pressure chemical vapor-deposited (LPCVD) silicon nitride (SiN) square membranes were investigated, with nominal tensile stresses of 200~MPa and 1100~MPa, respectively. The SiN was deposited on both sides of (100)-oriented single-crystal silicon wafers. 
The membranes were patterned by photolithography and reactive ion etching of the backside SiN layer and subsequently released by anisotropic KOH (40\,wt.\%) wet etching through the full wafer thickness.
The membrane thicknesses $h$ range from 10 to 300 nm, while the lateral dimensions $L$ are either 0.85 mm or 1 mm. The main properties of all the wafers used in this study are summarized in Table~\ref{tab:wafer_used}, while details about the fabrication of the 30 nm wafer are reported in the Supplementary Information.


\begin{table}[htbp]
    \centering
    \caption{List of SiN thin films studied.}
    \label{tab:wafer_used}
    \begin{tabular}{lccl}
        \toprule
        Label & h [nm] & $\sigma_{pre}$ [MPa] & Source \\
        \midrule
        10HS  & 10  & 1100 & Hahn-Schickard, Germany \\
        30LS  & 30  & 200  & QFactory, Denmark \\
        30HS  & 30  & 1100 & QFactory, Denmark \\
        50LS  & 50  & 200  & Hahn-Schickard, Germany \\
        300LS & 300 & 200  & Hahn-Schickard, Germany \\
        \bottomrule
    \end{tabular}
\end{table}
 
\subsection{UHV setup and thermal treatment protocol}\label{subsec:setup_thermal treatment}
 
\begin{figure*}
    \centering
    \includegraphics[width=\textwidth]{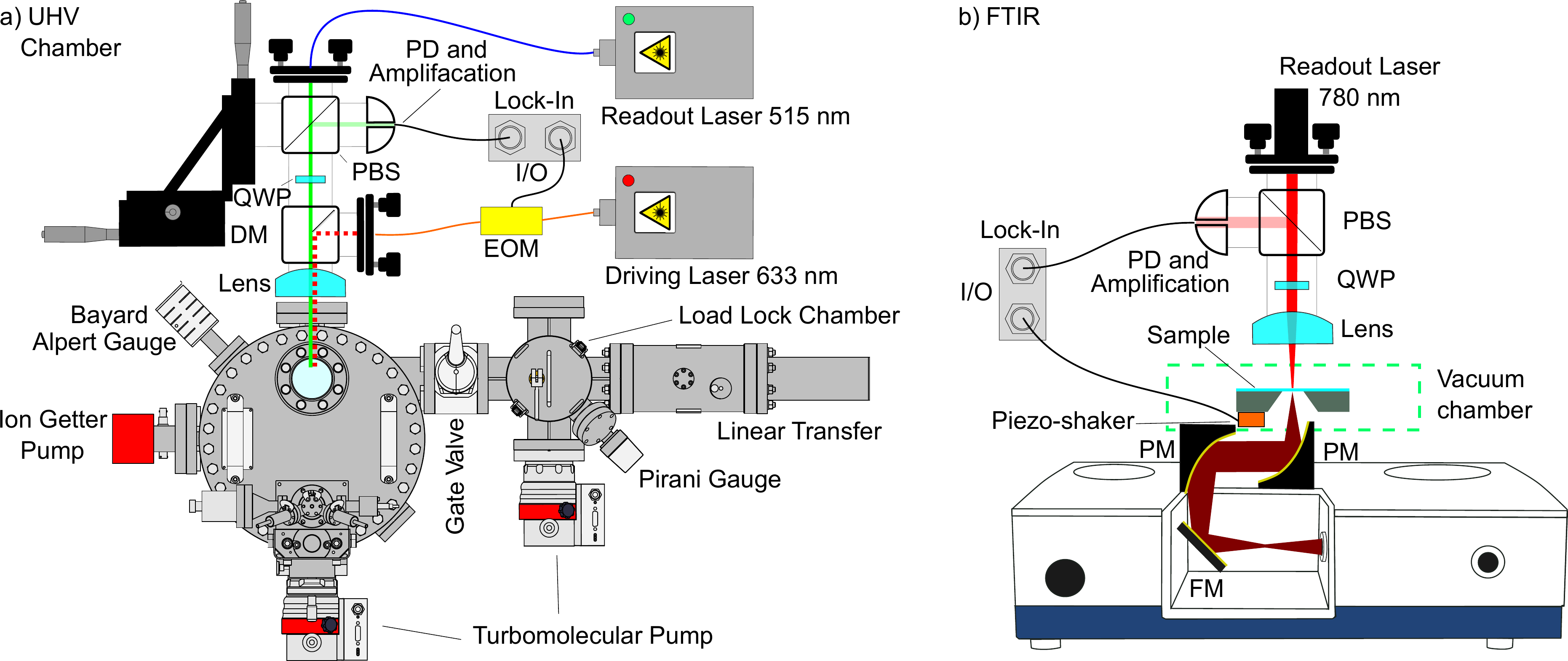}
    \caption{Experimental setups used in this work. (\textbf{a}) UHV and thermal-treatment setup. \emph{Top}: optical lever setup for the readout of SiN membrane resonators. A green 515~nm laser is used to detect the mechanical vibrations, while a red 633~nm laser is directed into the sample compartment by a dichroic mirror (DM) and drives the resonator through an electro-optic modulator (EOM). The reflected green light is collected and directed onto a four-quadrant photodetector (PD); the signal is preamplified, demodulated, and fed to a lock-in amplifier to close the feedback loop. A polarizing beam splitter (PBS), combined with a quarter-wave plate (QWP), is used to maximize the transmitted optical power.
 \emph{Bottom}: UHV system for thermal treatment, with load-lock chamber and the main components for vacuum measurement and control. (\textbf{b}) FTIR setup. Optical lever as in (\textbf{a}), with piezoelectric actuation of the membrane. The divergent FTIR beam is focused onto the backside of the SiN membrane inside a small vacuum chamber by two parabolic mirrors (PM), with a flat mirror (FM) redirecting the beam.}
    \label{fig:setups}
\end{figure*}

Thermal treatment is performed in ultra-high vacuum ($p < 5\cdot10^{-10}$ mbar) using a PREVAC HEAT3-PS heating stage with PID-controlled temperature ramping. The stage operates in either resistive or electron-bombardment mode: the resistive mode uses a 1-inch chuck and reaches \SI{1000}{\celsius}, the electron-bombardment mode a half-inch chuck reaching \SI{2000}{\celsius}. 

The ramp-up and ramp-down rates are set to \SI{40}{\celsius\per\minute}, and the hold time at the target temperature is fixed at 10 minutes if not otherwise stated. These values are a compromise: slower ramp rates produce no observable change in the results, while longer hold times degrade the quality factor and reduce the tensile stress at higher temperatures. The temperatures quoted throughout refer to the heating-stage thermocouple. Finite-element simulations of the membrane temperature profile (see Supplementary Information) indicate that the membrane edge closely follows the chuck temperature, whereas the center remains below it, with radiative
exchange partially compensating the deficit. The reported treatment
temperatures should therefore be understood as an upper bound for the
membrane center. The relevant temperature is the one at the locations where dissipation occurs, which, for membrane resonators, is weighted toward the strongly curved clamping region that closely follows the chuck temperature~\cite{shaniv2023direct}. The reported enhancements are conservative with respect to a spatially uniform treatment. Repeating the treatment at the same target temperature and ramp rate yields reproducible results, indicating that the process is governed by thermodynamic equilibrium rather than kinetic limitations (see Supplementary Information).

The samples are measured \textit{in situ} at room temperature using the setup in Fig.~\ref{fig:setups}a. Resonance modes are optically driven with a 633~nm laser (Toptica TopMode) modulated by an electro-optic modulator (Jenoptik AM635b), delivering up to 7~mW on the sample. Displacement is read out by an optical lever~\cite{putman_OLcomparison1992} at normal incidence, using a 515~nm laser (Toptica TopMode). Both readout and drive powers are adjusted by variable neutral-density filters. A polarizing beam splitter followed by a quarter-wave plate directs the reflected light onto a silicon four-quadrant photodetector (QP5.8-6 TO, First Sensor); the photocurrents are amplified by a transimpedance amplifier, subtracted by an instrumentation amplifier, and sent to a lock-in amplifier (Zurich Instruments MFLI or UHF). Mode frequencies are tracked with a phase-locked loop, and quality factors are measured by ring-down~\cite{schmid_fundamentals}: with the mode locked, the drive is switched off and the decay time constant is extracted from an exponential fit to the ring-down amplitude.

\subsection{Intrinsic quality factor determination}\label{subsec:qint_determination}
 

The intrinsic quality factor $Q_\mathrm{int}$ of a dissipation-diluted square membrane of side length $L$ and thickness $h$ is obtained from the measured $Q$ as~\cite{schmid_fundamentals}
\begin{equation}\label{eq:dissipation_dilution_Q}
        Q_\mathrm{int} = \left[\frac{(n^2+j^2)\,\pi^2}{12}\,\frac{E}{\sigma}\left(\frac{h}{L}\right)^2 + \frac{1}{\sqrt{3}}\sqrt{\frac{E}{\sigma}}\,\frac{h}{L}\right] Q,
\end{equation}
where $n$ and $j$ are the mode numbers, $E$ is the Young's modulus, and $\sigma$ the tensile stress. We neglect the weak frequency and stress dependence of $E$ and take $E = 250$~GPa throughout \cite{kanellopulos2024stress}. The stress is extracted from the square-membrane dispersion relation,
\begin{equation}\label{eq:resonance_freq}
    f_{n,j} = \frac{1}{2L}\sqrt{\frac{\sigma}{\rho}}\,\sqrt{n^2+j^2},
\end{equation}
by a linear regression of $f_{n,j}^2$ against $n^2+j^2$, whose slope $\sigma/(4\rho L^2)$ yields $\sigma$. The mass density is taken as $\rho = 3000~\mathrm{kg/m^3}$~\cite{muller2026devil}.

For each sample, we measure $Q$ for a large number of high-order modes ($n,j \gg 1$) using a custom Python routine. $Q_\mathrm{int}$ is extracted by fitting Eq.~\eqref{eq:dissipation_dilution_Q} to the upper envelope of the measured $Q$ values as a function of $n^2+j^2$, and the fitted envelope is taken as the intrinsic quality factor of the sample (Fig.~\ref{fig:Qenhancement}a). The envelope is used because $Q$ can scatter by order of magnitudes within a single sample: modes lying below the envelope are not limited by intrinsic material loss but carry additional, mode-dependent losses, which we attribute to acoustic radiation into the silicon frame \cite{Silvan2014_evidence_of_surf_loss, benga2026determination}. In extracting $Q_\mathrm{int}$, we neglect the weak frequency dependence of the loss modulus (see, e.g., Ref.~\cite{schmid_fundamentals}).


\subsection{Photothermal FTIR characterization}\label{subsec:ftir_methods}
 
   
FTIR characterization is performed by nanomechanical photothermal spectroscopy~\cite{jelena2026, west2023_photothermal}, with the setup shown in Fig.~\ref{fig:setups}b. The SiN membrane resonator is actuated piezoelectrically and read out with a 2~mW, 780~nm diode laser (Roithner GmbH). The broadband beam from the FTIR spectrometer (Bruker Vertex 70) is collimated and focused onto the membrane by a 1-inch gold-protected parabolic mirror. The mechanical mode is tracked by a phase-locked loop, and the resulting frequency-shift interferogram is fed back into the spectrometer, which performs a Fourier transform to recover the absorption spectrum. Details of the XPS characterization are given in the Supplementary Information.

\section{Results and Discussion}\label{sec:results}

\subsection{Thermal treatment in UHV}\label{subsec:q_enhancement}

\begin{figure*}
    \centering
    \includegraphics[width=0.7\textwidth]{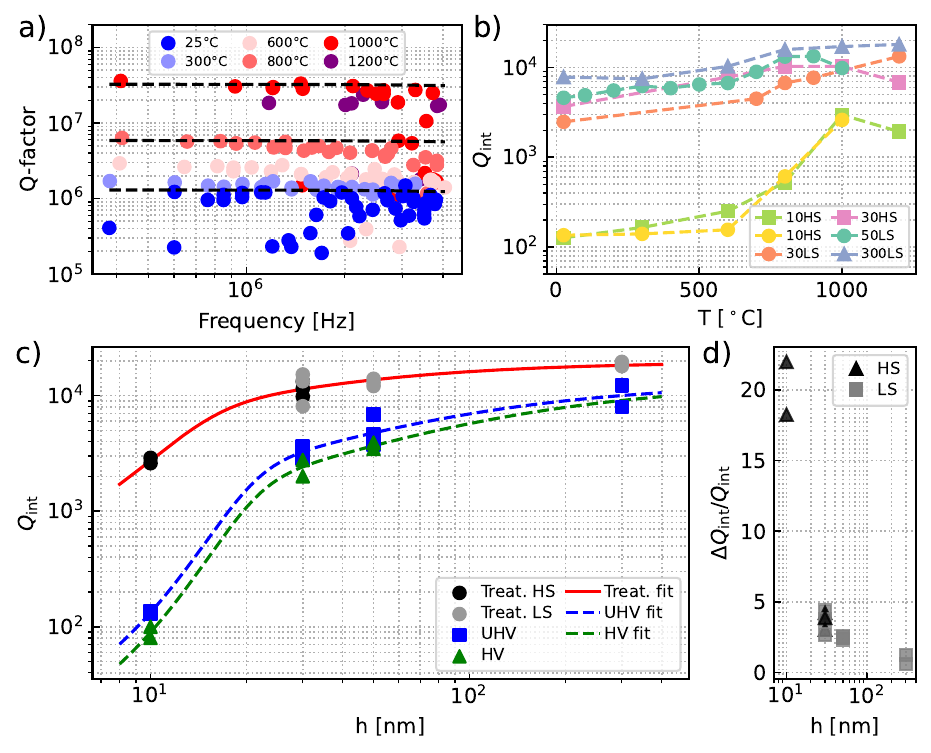}
    \caption{Quality factor enhancement by thermal treatment. (\textbf{a})~Quality factors of all measured modes of a 10HS sample versus frequency, before treatment ($25^\circ$C) and after successive 10-minute treatments up to $1200^\circ$C. The dashed black lines are the envelope fit obtained for the sample after 25, 600 and 1000$^\circ$C. (\textbf{b})~Intrinsic quality factor at room temperature after 10-minute treatments at increasing temperatures; each curve corresponds to one sample (Table~\ref{tab:wafer_used}). (\textbf{c})~Thickness dependence of $Q_\mathrm{int}$ measured initially in HV (green), then in UHV (blue), and after treatment at $1000^\circ$C (gray and black for LS and HS, respectively). Lines are fits of Eq.~\eqref{eq:Qint_surf_limited} with the parameters of Table~\ref{tab:diss-channels}. (\textbf{d})~Relative improvement of $Q_\mathrm{int}$ versus film thickness, reaching a factor of 20 for the 10HS membranes.}
    \label{fig:Qenhancement}
\end{figure*}

Fig.~\ref{fig:Qenhancement}a shows the quality factors of all measured modes of a 10~nm high-stress SiN membrane as a function of resonance frequency before treatment (25$^\circ$C) and after thermal treatment cycles performed at 300, 600, 800, 1000, and $1200^\circ$C. The values increase from approximately 1 million to over 30 million after treatment at 1000$^\circ$C, then they decrease to 20 million after 1200$^\circ$C. 

Fig.~\ref{fig:Qenhancement}b plots the extracted intrinsic quality factors (\ref{eq:dissipation_dilution_Q}) of all measured samples (Table~\ref{tab:wafer_used}) after thermal treatments with increasing temperatures up to $1200^\circ$C.  An overall monotonic increase of $Q_{\mathrm{int}}$ is observed up to 1000$^\circ$C, with a stronger increase after treatments $> 500^\circ$C, which is especially pronounced for the 10 nm membranes.

Fig.~\ref{fig:Qenhancement}c presents the intrinsic quality factors of all measured samples under three different conditions: i) initially in high vacuum (HV), ii) then in UHV, and iii) after thermal treatment at $1000^\circ$C. The experimental data are fitted with (\ref{eq:Qint_surf_limited}) and the fit parameters and their improvements are listed in Table~\ref{tab:diss-channels}. Moving from HV to UHV already improves the surface and excess-loss channels by a factor of 1.4 and 1.5, respectively. A major improvement across all three aforementioned loss channels (\ref{eq:Qint_surf_limited}) is achieved after the thermal treatment, with 8-fold and 50-fold improvements of the surface and excess-loss channels, respectively. Even the volume losses see a slight reduction of approximately 35$\%$. In total, the thermal treatment in UHV resulted in a more than 20-fold improvement of the intrinsic quality factor of the 10~nm SiN membranes, as shown in Fig.~\ref{fig:Qenhancement}d.

The improvement observed upon merely moving from HV to UHV, prior to any heating, is consistent with the partial removal of adsorbed molecular water. At the HV pressure of $5\cdot10^{-6}$~mbar, gas friction can be excluded (see Supplementary Information). Physisorbed water on hydroxylated silicon nitride is comparatively weakly bound (adsorption enthalpy $\approx0.5$~eV, as measured calorimetrically on crystalline Si$_3$N$_4$~\cite{fubini1989reactivity}), and its coverage is governed by the balance between the adsorption flux, proportional to the water partial pressure, and thermal desorption; in the low-coverage limit the equilibrium coverage scales linearly with pressure.
Reducing the pressure by roughly three orders of magnitude is therefore
expected to further deplete the weakly bound fraction of the admolecule layer,
consistent with calorimetric observations that reversibly adsorbed molecular water leaves Si$_3$N$_4$ surfaces upon evacuation at room temperature alone~\cite{Asay_2005_adH2O_SiO2}. Vacuum and thermal treatment thus act preferentially on different tiers of surface water: evacuation removes the weakly hydrogen-bonded molecular fraction, while heating is required to remove more strongly bound interfacial water and, ultimately, the chemically bound hydroxyl terminations themselves. We note that this picture is necessarily simplified: interfacial water is structured, and the first, more strongly bound (ice-like) layers desorb slowly and incompletely at room temperature, as water adsorbs in high energy defects on the surface. This was evidenced by the need  of heating to precondition samples for water adsorption--desorption under nitrogen purging\cite{jagerska2026water} or moderate vacuum ~\cite{fubini1989reactivity}.

\begin{table*}[t]
  \centering
  \caption{Fit parameters of the three dissipation channels (\ref{eq:Qint_surf_limited}) after each
    surface-treatment step, and their step-to-step improvement factors.
    $Q_\mathrm{S}$ and $\beta$ parameterize the same surface channel
    ($Q_\mathrm{S}=6\delta\beta$, with $\delta=0.4$\,nm \cite{benga2026determination}).}
  \label{tab:diss-channels}
  \begin{tabular}{l ccc c ccc}
    \toprule
    & \multicolumn{3}{c}{Fit value} & & \multicolumn{3}{c}{Improvement factor} \\
    \cmidrule(lr){2-4}\cmidrule(lr){6-8}
    Parameter & HV & UHV & $1000\,^{\circ}$C & &
      UHV & $1000\,^{\circ}$C & total \\
    \midrule
    $Q_\mathrm{V}^{\infty}$   & $12900^{\dagger}$& $12900 \pm 3400$& $19600 \pm 2900$& & ---& $1.5$& $1.5$\\
    \addlinespace[2pt]
    $Q_\mathrm{S}$& $270 \pm 24$& $360 \pm 80$& $2200 \pm 600$& & \multirow{2}{*}{$1.4$}& \multirow{2}{*}{$6.0$}& \multirow{2}{*}{$8.2$}\\
    $\beta$ (nm$^{-1}$)       & $111 \pm 10$ & $150 \pm 40$& $910 \pm 260$& &        &       &       \\
    \addlinespace[2pt]
    $Q_\mathrm{V}^{\mathrm{def}}$ & $8.1 \pm 0.9$ & $12.3 \pm 2.0$& $410 \pm 80$& & $1.5$& $33$& $50$\\
    \bottomrule
  \end{tabular}
  \\[2pt]
  {\footnotesize $^{\dagger}$Fixed to the UHV value in the fit.}
\end{table*}

Our surface-loss coefficient $\beta$ measured in HV exceeds the literature value of $\beta \approx 60$~nm$^{-1}$, while the extracted $Q_\mathrm{V}^{\infty}=12\,900$ is roughly half the previously reported value of $28\,000$, i.e., the volume loss is about twice as large~\cite{Silvan2014_evidence_of_surf_loss}. The low  surface losses indicate the high quality and cleanliness of our SiN thin films and may also be related to differences in sample storage conditions (see Supplementary Information). The larger volume loss suggests that previous values extracted from a meta-analysis of literature data likely overestimated $Q_\mathrm{V}^{\infty}$, which was also confirmed by \cite{benga2026determination}.

The 50\% increase in $Q_\mathrm{V}^{\infty}$ after thermal treatment (see Table~\ref{tab:diss-channels}) cannot be fully explained by changes in the ratio $E/\rho$, which would affect the extraction of $Q_{\mathrm{int}}$ (\ref{eq:dissipation_dilution_Q}). Previous studies have reported variations in these quantities of less than approximately 10\% after thermal treatment at $1100^\circ$C~\cite{jiang_anneal_SiN_2016} , with a smaller expected change for the ratio $E/\rho$.

The low $Q_\mathrm{int}$ of the 10~nm samples reflects the degraded morphology of ultrathin films and is captured by the excess volume loss term in (\ref{eq:Qint_surf_limited}). The CVD growth of SiN on silicon oxide (e.g., the native oxide of the Si substrate) proceeds by island formation, with the islands merging only at a film thickness of 2--3~nm \cite{copel1999nucleation}; chemical \cite{weinberg1990ultrathin} and electrical \cite{ando1991ultrathin} permeability tests find a critical thickness of 6--7~nm before SiN films become impermeable. Consistently, the refractive index of the 10~nm films, $n=1.75$, is significantly lower than the $n\approx2.0$ of the 30~nm films (see Supplementary Information); accounting for the 1--2~nm surface oxide with $n=1.4$ \cite{giesriegl2026hf,Nik2017_plasma_ox} yields $n\approx1.9$ for the SiN itself, indicating excess oxide content and/or a reduced mass density in the defective nucleation region. Based on this, the characteristic thickness in (\ref{eq:Qint_surf_limited}) was fixed at $h_\mathrm{def}=4$~nm.
Notably, this defective nucleation layer responds even more strongly to the thermal treatment than the outer surfaces (50-fold versus 8-fold, Table~\ref{tab:diss-channels}); we return to this observation in Sec.~\ref{subsec:air_exposure}.

For the thickness of 30~nm, no significant difference in intrinsic dissipation is observed between low-stress and high-stress samples. This justifies treating both datasets together and suggests that they share a common dissipation mechanism.

\begin{figure*}
    \centering
    \includegraphics[width=0.7\textwidth] {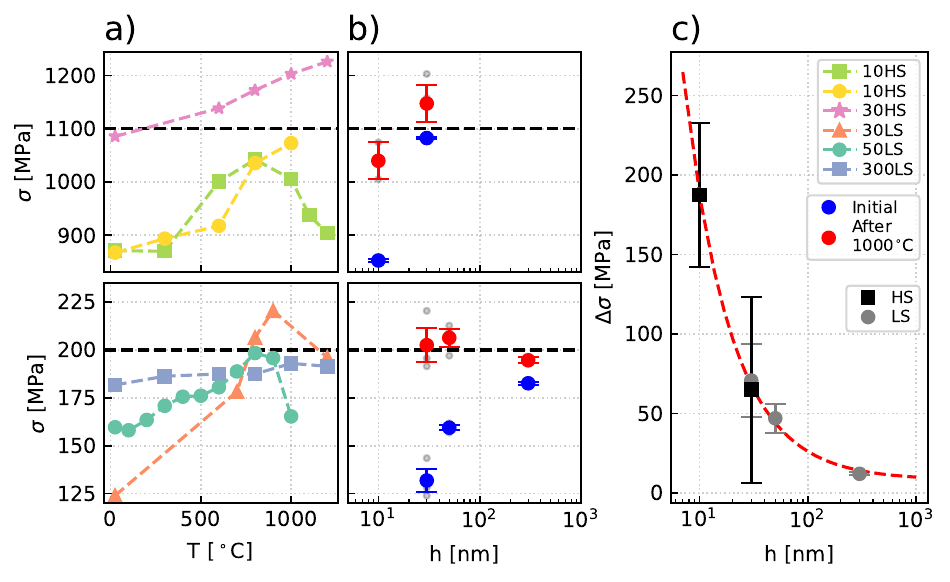}
    \caption{Stress modification by thermal treatment. (\textbf{a})~Room-temperature tensile stress after treatment at increasing temperatures for the 1100~MPa (\emph{top}) and 200~MPa samples (\emph{bottom}). (\textbf{b})~Stress before (blue) and after (red) treatment at $1000^\circ$C versus film thickness. (\textbf{c})~Stress change $\Delta\sigma$ after the $1000^\circ$C treatment versus thickness. The dashed line is a fit to Eq. \ref{eq:surface_stress_scaling}.}
    \label{fig:stress_change}
\end{figure*}

The thermal treatment also affects the tensile stress. Fig.~\ref{fig:stress_change}a shows the tensile stress as a function of heating temperature. The thermal treatment increases the tensile stress in all samples up to approximately 800--1000$^\circ$C. At higher temperatures, a reduction in stress is observed for all the membranes except the 50~nm high-stress sample. Together with the concurrent decrease of $Q_\mathrm{int}$ after treatment at $1200^\circ$C (Fig.~\ref{fig:Qenhancement}a,b), this suggests the onset of structural relaxation of the amorphous network~\cite{Smeys_viscus_1995,jiang_anneal_SiN_2016,Noskov88_corr_stress_H}, whose detrimental effect at these temperatures outweighs the enhancement observed at lower temperatures.

Fig.~\ref{fig:stress_change}b shows the stress before and after thermal
treatment at \SI{1000}{\celsius} as a function of SiN thickness. The
corresponding stress change $\Delta\sigma$ of all samples, high-stress and
low-stress, is summarized in Fig.~\ref{fig:stress_change}c. The increase in
stress is largely independent of the initial prestress but scales
approximately as $1/h$, identifying a surface-localized origin. Several surface contributions
could account for a change of this magnitude: the native oxide and oxynitride
layer, whose growth or removal is known to alter the stress of thin SiN
films~\cite{Nik2017_plasma_ox}, the desorption of water and carbon
adsorbates~\cite{stgelais_surfenergy2025}, or a modification of the chemical
termination of the surface itself~\cite{stephan2002}. We note, however, that the stress increase far above the temperatures at which physisorbed water desorbs from
Si$_3$N$_4$ surfaces~\cite{fubini1989reactivity}, which already disfavors simple adsorbate desorption as the origin of the high-temperature response. Physisorbed water and hydrocarbons are nonetheless expected to contribute to both $Q_\mathrm{int}$ and $\sigma$ at room temperature and to the weak response observed below \SI{500}{\celsius}, and their variable loading plausibly contributes to the sample-to-sample scatter commonly reported for as-received SiN resonators \cite{Silvan2014_evidence_of_surf_loss}. 

Finally, we consider whether the stress increase itself, through enhanced
dissipation dilution, could mimic a reduction of the intrinsic loss. Since
the surface reaction is thermally activated, the temperature inhomogeneity
across the membrane (see Supplementary Information) additionally implies a
spatially non-uniform stress change, of which the extracted stress is a
modal average. Both effects are, however, bounded by the relative stress
change --- below 10\% for the high-stress samples --- and cannot account
for the twentyfold enhancement of $Q_\mathrm{int}$.
The thermal-treatment results in UHV establish that the reduction in
intrinsic dissipation is mainly related to surface processes, together with
effects arising from the degraded morphology of the ultrathin films, with
maximum improvements achieved at 1000$^\circ$C. A surface-bound origin of
the observed changes is independently corroborated by the thickness
dependence of the increase in tensile stress. 

The mechanical data alone, however, do not identify which property of the surface is modified by the
treatment. The surfaces of SiN membranes carry a native oxide/oxynitride
layer, whose stress state could respond to thermal
treatment~\cite{Nik2017_plasma_ox}, and such oxidized surfaces are known to
be a dominant source of dissipation in nanomechanical resonators, with a
friction that depends on the structure and formation conditions of the
layer rather than on its amount
alone~\cite{tao_permanent_reduction_dissipation_2015}. The treatment could
therefore act on the amount of surface oxide, on its internal structure, or
on its chemical termination. To discriminate between these scenarios, we
characterize the chemical composition of the membrane surfaces before and
after thermal treatment with FTIR and XPS.

\subsection{Surface composition: FTIR and XPS}\label{subsec:compositional_analysis}

\begin{figure*}
\centering
    \includegraphics[width=0.8\textwidth] {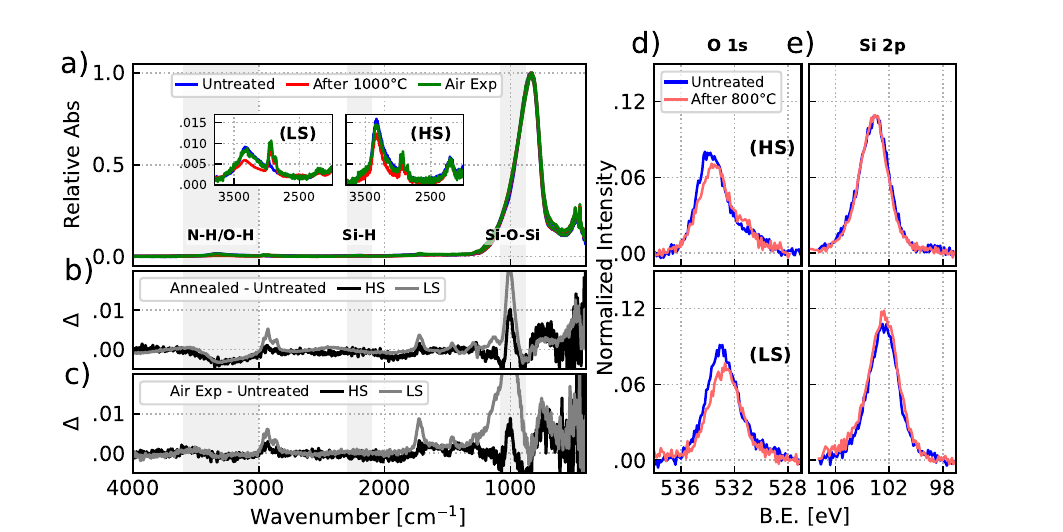}
    \caption{Surface composition in different conditions. (\textbf{a})~Photothermal FTIR spectra of high-stress (HS) and low-stress (LS) 30~nm SiN membranes before treatment, after treatment at $1000^\circ$C, and after subsequent air exposure, normalized to the Si--N peak at 850--900~cm$^{-1}$. Insets: Si--H stretching mode at 2200~cm$^{-1}$ and N--H/O--H stretching region at 3000--3600~cm$^{-1}$. (\textbf{b},\textbf{c})~Difference spectra relative to the untreated state, after treatment (b) and after subsequent air exposure (c). The negative band at 3300~cm$^{-1}$ indicates the loss of N--H/O--H terminations; the positive peak at 1010~cm$^{-1}$ the growth of Si--O--Si bonds; features at 1715~cm$^{-1}$ and 2800--3000~cm$^{-1}$ stem from carbon contamination acquired during air transfer. (\textbf{d})~\textit{In situ} XPS O~1s and Si 2p spectra of HS (\emph{top}) and LS (\emph{bottom}) 30~nm membranes before and after treatment at $800^\circ$C for 10 minutes.}
    \label{fig:FTIR_XPS}
\end{figure*}

Fig.~\ref{fig:FTIR_XPS}a presents photothermal FTIR spectra of 30HS and 30LS at three stages: before thermal treatment, after treatment at $1000^\circ$C, and after subsequent air exposure for two weeks, normalized to the dominant Si--N peak at 850--900~cm$^{-1}$. The insets show the Si--H stretching mode at 2200~cm$^{-1}$, barely visible for LS, and the N--H/O--H stretching region at 3000--3600~cm$^{-1}$.

The difference spectra (Fig.~\ref{fig:FTIR_XPS}b\&c) reveal two systematic changes upon thermal treatment: (i)~a negative band around 3300~cm$^{-1}$, indicating a loss of N--H/O--H groups that is almost fully reversed upon air exposure, and (ii)~a positive band near 1010~cm$^{-1}$, indicating a growth of Si--O--Si content associated with silicon oxynitride \cite{sanchez1998characterization}, consistent with the assignment in \cite{giesriegl2026hf,giesriegl2019}. In contrast, the absence of a differential signal at 2200~cm$^{-1}$ shows that the Si--H content remains unchanged within the detection limit. 
Additional positive features at 1715~cm$^{-1}$ (carbonyl stretching) and at
2800--3000~cm$^{-1}$ (C--H stretching) are attributed to carbon
contamination acquired during the $\sim$10-minute air transfer to the FTIR
setup. 

Two observations locate the change at the surface rather than in the bulk. First, the unchanged Si--H signal shows that the bulk hydrogen content is not affected, in contrast to treatments of several hours at \SI{1100}{\celsius}~\cite{jiang_anneal_SiN_2016,Mittal_H_2024}. Second, the N--H/Si--H peak ratio of the HS samples decreases from approximately 3 to 2 upon treatment, the latter matching values reported for 500--800~nm-thick films, in which surface contributions are negligible~\cite{Mittal_H_2024}. The removed species are therefore surface O--H and -NH$_2$ terminations, with the reduction dominated by O--H given the  negligible change of N 1s in the XPS data (see Supplementary Information).

Finally, the O--H signal is almost completely restored for both LS and HS samples after the samples have been re-exposed to air (Fig.~\ref{fig:FTIR_XPS}c). 

In summary, the FTIR data show a treatment-induced reduction in surface O--H  terminations, accompanied by an increase in Si--O--Si bonds and an unchanged bulk hydrogen content --- the spectroscopic fingerprint of thermally activated silanol condensation.

Surface-sensitive analysis was performed with \textit{in situ} XPS before and after thermal treatment at $800^\circ$C for 10 minutes (Fig.~\ref{fig:FTIR_XPS}d\&e); the treatment temperature was limited by the temperature gauge of the XPS setup. Nevertheless, $800^\circ$C is sufficient to induce substantial changes in the mechanical properties of the 30~nm samples (Figs.~\ref{fig:Qenhancement}b and~\ref{fig:stress_change}a). The XPS spectra reveal a modest reduction of the O~1s intensity, while the changes in the Si~2p are negligible. All variations are close to the instrument's sensitivity limit, indicating that the thermal treatment does not significantly alter the film's chemical composition. A second set of measurements showing the same behavior is reported in the Supplementary Information.

The XPS results thus rule out a removal or substantial modification of the oxide/oxynitride surface layer. The modest oxygen reduction is instead consistent with the loss of a sub-monolayer of hydroxyl groups, in line with the O--H reduction observed by FTIR. Together, the two techniques identify the treatment-induced change as a modification of the surface \emph{termination} within an intact oxide layer: thermally activated silanol condensation~\cite{zhuravlev2000, Comas2016_SiO2_surf_model},
\begin{equation}
\mathrm{SiOH} + \mathrm{SiOH} \rightleftharpoons  \mathrm{SiOSi} + \mathrm{H_2O(g)},
\label{eq:condensation_reaction}
\end{equation}
where elevated temperature and vacuum drive the equilibrium to the right.

This assignment is supported by calorimetric studies of water on crystalline Si$_3$N$_4$ surfaces~\cite{fubini1989reactivity} and on silica~\cite{bolis1991hydrophilic}: thermal treatment in vacuum at $400^\circ$C removes only weakly hydrogen-bonded molecular water ($-\Delta H \simeq 50$~kJ/mol $\approx 0.5$~eV), leaving OH and NH terminations intact, whereas at higher temperatures these groups progressively condense with release of chemically bound water, with desorption onsets reported already at 500--600$^\circ$C and condensation continuing up to 1000$^\circ$C~\cite{fubini1989reactivity,zhuravlev2000}  --- matching the temperature range in which we observe the steep increase in $Q_\mathrm{int}$ and tensile stress. The same study finds the condensed surface to be highly reactive toward
dissociative re-adsorption of water ($-\Delta H > 50$~kJ/mol), which
restores the OH/NH termination, consistent with the recovery of the
3300~cm$^{-1}$ band upon air exposure. At the same time, the loss of
hydroxyl groups reduces the affinity of the surface for molecular
water~\cite{bolis1991hydrophilic}, favoring instead the adsorption of
airborne organic species~\cite{wu2014,schlangen1995}. Together with the
known accumulation of adventitious carbon on SiN surfaces upon air
exposure~\cite{stgelais_surfenergy2025}, this explains the enhanced carbon
contamination we observe after treatment, consistent with an altered
surface affinity~\cite{giesriegl2026hf}. On our oxide-terminated surfaces, the condensation product is the siloxane bridge observed at 1010~cm$^{-1}$, rather than the Si--N bridges formed on pristine, nitride-terminated Si$_3$N$_4$ surfaces, where sparse, isolated silanols condense with neighboring N--H groups instead \cite{fubini1989reactivity}. 

Using the FTIR spectra and the absorption cross sections $x_{\mathrm{SiN}}$ and $x_{\mathrm{OH}}$ of the Si--N and O--H stretching modes~\cite{digregorio2020infrared,nagasawa2021absolute}, the surface hydroxyl density is estimated as
\begin{equation}
    n_\mathrm{OH} = n_{\mathrm{SiN,3D}} \, h \, \frac{x_{\mathrm{SiN}}}{x_{\mathrm{OH}}} \, \frac{A_{\mathrm{OH}}}{A_{\mathrm{SiN}}},
\end{equation}
where $A$ denotes the integrated peak area. For the LS samples, this yields an initial coverage of approximately $2.5~\mathrm{nm}^{-2}$, in agreement with typical values for SiO$_2$ surfaces~\cite{Mueller2003_Sio2densitiy}, and a reduction of approximately $0.9~\mathrm{nm}^{-2}$ upon thermal treatment. The residual density of -OH is likely overestimated, as partial rehydroxylation during the air transfer to the FTIR setup cannot be ruled out.

\subsection{Reversibility upon air exposure}\label{subsec:air_exposure}

\begin{figure*}
\centering
    \includegraphics[width=0.7\textwidth ] {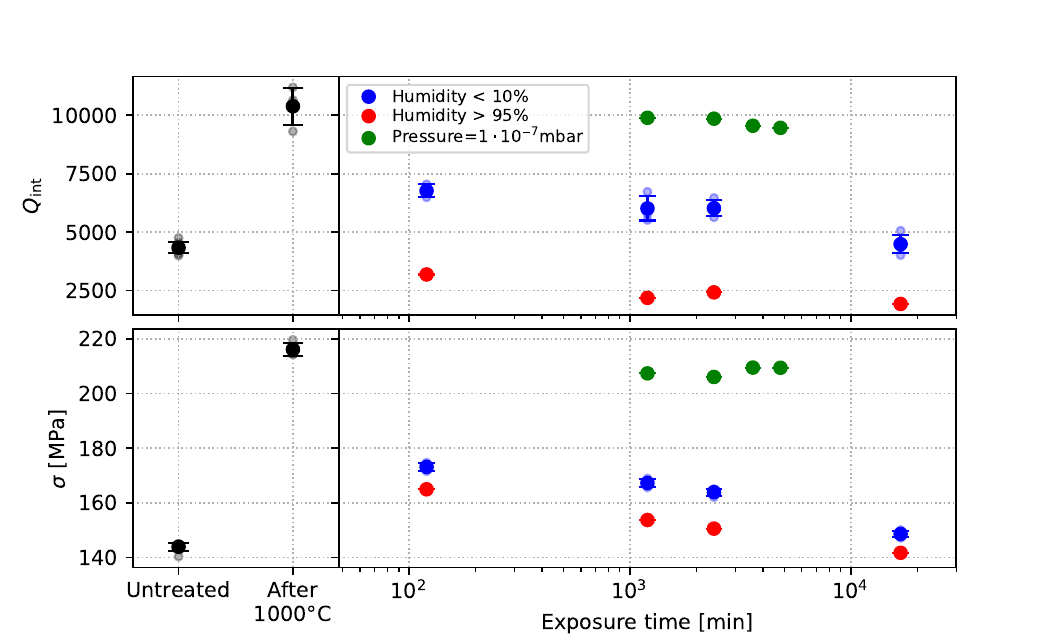}
    \caption{Reversibility of the treatment-induced enhancement in 30LS samples. \emph{Left}: $Q_\mathrm{int}$ and tensile stress for untreated samples and after treatment at \SI{1000}{\celsius} for 20 minutes (mean over all samples; gray points are individual samples). \emph{Right}: evolution during storage at room temperature in high vacuum ($10^{-7}$~mbar, green), dry atmosphere (relative humidity $<10\%$, blue), and humid atmosphere (relative humidity $>95\%$, red). Error bars denote the spread over samples stored in the same environment.}
    \label{fig:reversibility_air_exp}
\end{figure*}

The silanol-condensation equilibrium (Eq.~\ref{eq:condensation_reaction}) makes a direct prediction: exposure to water vapor should reverse the treatment-induced changes at a rate that increases with the water vapor partial pressure, whereas storage in vacuum should preserve them. We test this hypothesis by storing thermally treated samples under three controlled conditions.

Five 30LS samples were treated at \SI{1000}{\celsius} for 20 minutes and subsequently stored at room temperature in either a dry environment (silica-gel desiccant; relative humidity - RH - below the 10\% sensitivity limit of the hygrometer) or a humid environment (sealed container partially filled with liquid water; relative humidity above 95\%). At \SI{25}{\celsius}, these conditions correspond to water vapor partial pressures of at most 1.3~mbar and approximately 13 mbar, respectively. The transfer from the vacuum to the ambient did not affect the measurements significantly due to the relatively short amount of time required (3 minutes) and intermediate humidity values in the lab.  In a separate experiment, an additional sample was treated under identical conditions and stored in the load-lock chamber at approximately $10^{-7}$~mbar.

Fig.~\ref{fig:reversibility_air_exp} shows the evolution of $Q_\mathrm{int}$ and $\sigma$ with exposure time, and the three storage conditions confirm the hypothesis. The vacuum-stored sample retains the post-treatment values of both quantities over the full observation period, ruling out spontaneous relaxation of the treatment effect. In contrast, both quantities decrease progressively at atmospheric pressure, and faster in the humid than in the dry environment: after roughly two hours, $Q_\mathrm{int}$ in the humid environment has dropped below its pre-treatment level, while the dry samples reach pre-treatment values only after more than one week. The >95\% RH environment exceeds typical ambient humidity, plausibly driving OH coverage and adsorbed water beyond the standard equilibrium. The tensile stress follows the same trend, relaxing from 216 MPa toward the initial 144 MPa. The systematically higher values observed in dry environment are consistent with the condensation equilibrium described by Eq.~\ref{eq:condensation_reaction}. 
Consistently, a subsequent treatment at 150$^\circ$C of the humid-stored
samples, sufficient to desorb molecular water~\cite{jagerska2026water}, restored
neither $Q_\mathrm{int}$ nor the stress (see Supplementary Information),
showing that the degradation is caused by rehydroxylation of the surface
rather than by adsorbed water itself. 

The combined observations of the previous sections point towards silanol condensation as the origin of the thermal-treatment-induced changes. First, the $Q_\mathrm{int}$ and $\sigma$ increase develops predominantly above 500 $^\circ$C, with the largest gains between 800 and 1000 $^\circ$C, incompatible with the desorption of molecular water, which is complete at far lower temperatures~\cite{fubini1989reactivity}, but coinciding with the temperature
range in which silanols condense~\cite{zhuravlev2000}. Second, XPS shows only a modest reduction of the surface oxygen content upon
annealing at 800$^\circ$C, consistent with the sub-monolayer hydroxyl loss
expected from the condensation reaction (Eq.\ref{eq:condensation_reaction})
and corroborated by the concomitant decrease of the O--H and increase of the
Si--O--Si signals in FTIR. The oxide layer itself remains intact, as
expected below the temperatures reported for thermal desorption of the
native oxide from silicon surfaces~\cite{SurfQ_Si_Yang2000,Yang2002}. Since
$Q_\mathrm{int}$ has already improved severalfold at this temperature
(Fig.~2b), the dissipation is decoupled from the amount of surface oxide,
ruling out oxide growth or removal~\cite{Nik2017_plasma_ox} as the origin of
the enhancement. Third, both stress
and quality factor recover under exposure to humid air at room temperature.
A structural relaxation of the oxide, such as densification at high
temperature, would be irreversible under these conditions; the hydrolysis of
strained siloxane bridges back into silanols, in contrast, is a known
room-temperature reaction.  The condensation of silanol groups into siloxane
bridges contracts the surface
network and adds a tensile surface-stress contribution~\cite{cammarata1994_surf_stress}:
\begin{equation}
\Delta \sigma = \frac{2 \Delta \gamma}{h}+k,
\label{eq:surface_stress_scaling}
\end{equation}
where $\Delta \gamma$ is the surface stress change induced by thermal treatment and the factor of two accounts for the two free surfaces of the membrane, while $k$  takes into account contribution from the bulk. From the fit of Fig. \ref{fig:stress_change}c we extract
\begin{equation*}
\Delta \gamma = 0.97 \pm 0.16~\mathrm{N/m}, \qquad k = 5.6 \pm 1.5~\mathrm{MPa}.
\end{equation*}
The calculated magnitude of  $ \Delta\gamma $ is consistent with
this picture: it corresponds to the chemisorption scale, comparable to the
surface stress reported for water chemisorption on
SiN~\cite{stephan2002,stgelais_surfenergy2025}, as expected for the formation or removal of a sub-monolayer density ($\sim 1 ~{nm}^{-2}$) of covalent surface bonds. The same surface reaction thus provides a common, quantitatively consistent origin for both the stress increase and the enhancement of the intrinsic quality factor.

Adsorbed hydrocarbons can be excluded as the origin of the enhancement on the
same quantitative grounds. Adventitious carbon accumulates on all samples
exposed to air, so the degradation kinetics alone do not discriminate between mechanisms. The tensile stress does. Physisorption energies lie well below those of covalent bonds, and the adsorption-induced surface stress accessible to physisorbed layers is orders of magnitude below the $\Delta\gamma \approx 1$~N/m extracted here~\cite{stgelais_surfenergy2025}; hydrocarbon adsorption furthermore offers no mechanism for a reversible stress change that tracks $Q_\mathrm{int}$ across all three storage environments. In the forward direction, adventitious carbon is largely removed by vacuum annealing at a few hundred degrees Celsius~\cite{kerber2012study}, well below the \SI{500}{\celsius} onset of the enhancement reported here. Residual contributions from carbon or from oxygen incorporation cannot be excluded \textit{a priori}~\cite{tao_permanent_reduction_dissipation_2015,Wang2004}, but neither reproduces the correlated evolution of $Q_\mathrm{int}$ and $\sigma$.

This resolves an ambiguity inherent to experiments in which the surface
oxide is removed, thermally~\cite{SurfQ_Si_Yang2000,Yang2002} or
chemically~\cite{tao_permanent_reduction_dissipation_2015}, and regrows
upon air exposure: there, the amount of oxide and the state of its
termination change simultaneously. In our case, the oxide amount remains
constant while the termination is modulated thermally and, reversibly, by
humidity. The dissipation follows the
termination, identifying the hydroxyl groups, rather than the
oxide--nitride interface or the oxide volume, as the dominant lossy
element, consistent with the earlier observation that same-thickness
oxides with different atomic arrangement exhibit markedly different
friction~\cite{tao_permanent_reduction_dissipation_2015}.

The same full reversibility is observed for the 10~nm membranes, whose dissipation is dominated by the excess-loss channel: upon air exposure, they return completely to their pre-treatment values of both $Q_\mathrm{int}$ and stress (see Supplementary Information). This observation points to the nature of the excess loss. Since the nucleation layer forms at the wafer-side interface, it constitutes the exposed backside surface of the released membrane, and its defective, oxide-rich network can present hydroxylated internal surfaces accessible to the gas phase, which is consistent with both the exceptionally strong response of this channel to the treatment (50-fold, Table~\ref{tab:diss-channels}) and its complete rehydroxylation by ambient water vapor, which would not be expected for a sealed, buried defect layer. In thicker films, no comparable excess loss is observed, suggesting that this defective interfacial region does not persist unaltered but is largely eliminated during the continued deposition, which is in line with the reported impermeability threshold of 6--7~nm~\cite{weinberg1990ultrathin,ando1991ultrathin} and the increase of the refractive index at 30~nm (Supplementary Information). The exact morphology of this region and the pathways by which water accesses it remain to be established. Within this picture, the excess volume loss of ultrathin films is not a bulk property but surface-type dissipation with an enhanced effective surface area, governed by the same silanol chemistry as the outer faces, such that all three loss channels of Eq.~\eqref{eq:Qint_surf_limited} respond to a single underlying surface process.

\section{Conclusion and Outlook}\label{sec:conclusion}
Thermal treatment in ultrahigh vacuum reduces the intrinsic dissipation of dissipation-diluted SiN membrane resonators by up to a factor of 20, raising the surface-loss coefficient $\beta$ eightfold, from approximately 110 to 910~nm$^{-1}$, and the excess loss of ultrathin films 50-fold. FTIR, \textit{in situ} XPS, and the humidity-dependent reversibility of both $Q_\mathrm{int}$ and tensile stress identify thermally activated silanol condensation as the underlying process: the dissipation follows the hydroxyl termination of an otherwise intact oxide layer. Surface loss, treated for a decade as a fixed material property, is thus a chemically tunable parameter.
 
Two questions remain open. The modest response below $\sim$\SI{500}{\celsius}, where hydrogen-bonded vicinal silanols are already expected to condense~\cite{zhuravlev2000}, indicates that the loss is not simply proportional to the total hydroxyl coverage; its dependence on hydroxyl configuration and on frequency is unresolved. And since rehydroxylation by ambient water vapor reverses the enhancement within days, practical use outside vacuum environments requires
chemical passivation of the condensed surface~\cite{Henry2003,Henry2004,tao_permanent_reduction_dissipation_2015,giesriegl2026hf}.
 
Because $Q_\mathrm{int}$ enters the dissipation-diluted quality factor as a geometry-independent material prefactor, the demonstrated enhancement transfers directly to strain-engineered resonators, where soft-clamped defect modes~\cite{Tsaturyan2017_PnC} would multiply the same $Q_\mathrm{int}$ by dilution factors far exceeding those of the fundamental membrane modes studied here. Realizing this potential requires heating the region where the defect-mode dissipation is localized~\cite{shaniv2023direct}, which chuck heating reaches least: in our setup, the membrane center remains several hundred degrees below the frame. Targeted photothermal treatment, illuminating the Si--N absorption band around 900~cm$^{-1}$, would deliver the heat directly to the defect and enable an \textit{in situ} treatment of structured resonators for which oven-scale processing of the full chip is undesirable.

\begin{acknowledgments} \label{sec:acknowledgements}
The authors wish to thank Sophia Schneider, Patrick Meyer, and Michael Buchholz for their support with sample fabrication; Eric Langman for the 30 nm wafers; and Zifan Che for the fabrication of the 10 nm samples. We also thank Johannes Schalko and Gernot Fleckl for their assistance with the vacuum equipment, and Jannis Berger for his help with workshop manufacturing. We further acknowledge Johannes Steurer and Hajrudin Be\v{s}i\'c for their contributions to the realization of the preamplifier for the optical lever, and Johannes Hiesberger for his help in setting up the FTIR measurements and for fruitful discussions about the optical lever. 
Finally, we gratefully acknowledge Jürgen Smoliner, Alois Lugstein, and Masiar Sistani from the Institute of Solid State Electronics at TU Wien for granting access to their FTIR spectrometer. This project received funding from the Austrian Science Fund (FWF) under project 10.55776/I6086 and from the European Union's Horizon Europe research and innovation programme under the Marie Skłodowska-Curie grant agreement No. 101152469.
\end{acknowledgments}

\bibliography{biblio}

\end{document}


\begin{frontmatter}
\end{frontmatter}
\newpage

\section{30nm-wafer fabrication}
The details of fabrication of the 30nm wafer high and low stress are reported in the following. The high-stress (1100 MPa) silicon nitride film was deposited over 45 minutes at a pressure of 200 mTorr and an average temperature of $800^\circ$C, using dichlorosilane and ammonia precursor flow rates of 20 and 80 sccm, respectively. Similarly, the low-stress (200 MPa) film was deposited over 30 minutes at a pressure of 150 mTorr and an average temperature of $830^\circ$C, using dichlorosilane and ammonia flow rates of 200 and 50 sccm, respectively.

\section{Repeated Annealing}
\begin{figure}
    \centering

    \subfloat[]{
        \includegraphics[width=0.46\textwidth]{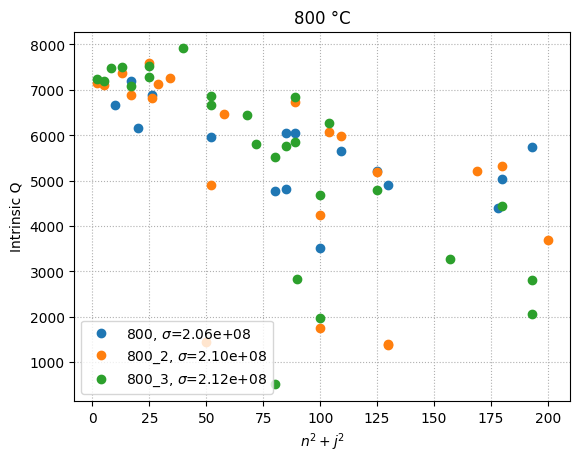}
        \label{fig:UHV10_repeated_ann800C_suppl}
    }
    \hfill
    \subfloat[]{
        \includegraphics[width=0.46\textwidth]{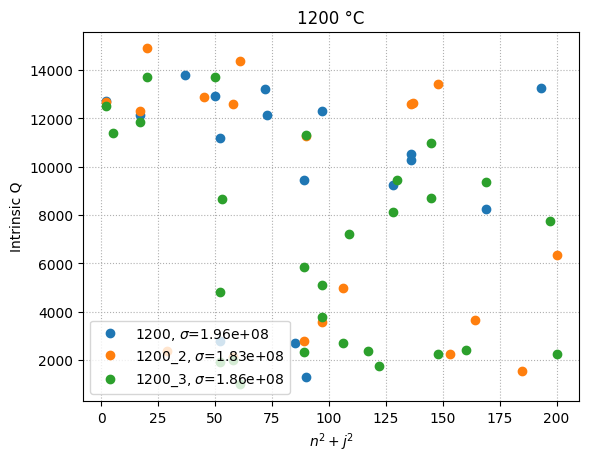}
        \label{fig:UHV10_repeated_ann1200C_suppl}
    }

    \caption{
    Low-stress 30 nm sample UHV10. Intrinsic \(Q\) measured after three different annealing cycles for 10 min at a heating rate of \(40^\circ\mathrm{C}/\mathrm{min}\) at \(800^\circ\mathrm{C}\) (a) and \(1200^\circ\mathrm{C}\) (b).
    }
    \label{fig:UHV10_repeated_ann_supp}
\end{figure}

 Annealing cycles with the same ramp velocity (40$^{\circ}$C/min), target temperature (800$^{\circ}$C, 1200$^{\circ}$C), and annealing time (10 min)  were repeated three times for a 30nm-low-stress sample. The results are presented in Fig.~\ref{fig:UHV10_repeated_ann_supp}, showing that $Q_{\mathrm{int}}$ and $\sigma$ are essentially unaffected. These results suggest that the bottle-neck of thermal treatments in these conditions is thermodynamics more than kinetics.

\section{Steady-state simulation of temperature profile}

 \begin{figure}[t]
    \centering
    \includegraphics[width=7cm]{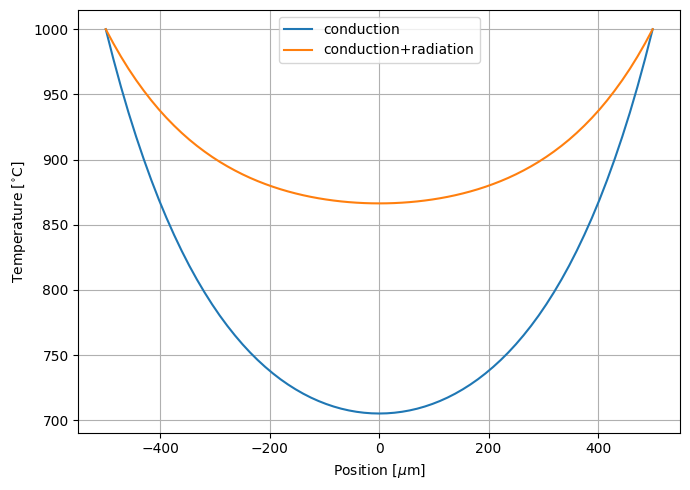}
\caption{Temperature profile of a 30nm-thick-SiN membrane with $L=1$\,mm simulated considering only thermal conduction (blue) and conduction including radiative contributions (orange). }
    \label{fig:comsol_sim}
\end{figure}

The temperature distribution during chuck annealing was calculated using the Heat Transfer in Solids module in COMSOL Multiphysics under steady-state conditions. The model solves the heat conduction equation:

\begin{equation}
    -\nabla \cdot (k_T\nabla T)=0
\end{equation}

where $k_T$ is the thermal conductivity and $T$ is the local temperature. The transient and convective terms are neglected, since a stationary study was performed and no material motion is considered.

The surrounding silicon substrate was not included in the model. The membrane is not in direct thermal contact with the chuck; instead, the conductive heat transfer from the chuck to the membrane occurs through the silicon frame. The perimeter of the membrane was assumed to be thermally anchored to the chuck temperature $T=T_{chuck}$. This boundary condition represents the silicon frame acting as a thermal reservoir during annealing.

Radiative heat exchange from the exposed membrane surfaces was included using the Surface-to-Ambient Radiation boundary condition. The top surface was assumed to exchange radiation with the chamber at room temperature, while the bottom surface was assumed to exchange radiation with the heated chuck.
The radiative heat flux was calculated according to the Stefan–Boltzmann law:

\begin{equation}
   q_{rad}=\epsilon_{eff}\sigma_{SB}
\left(T^4-T_{amb}^4\right) 
\end{equation}

where $\sigma_{SB}$ is the Stefan–Boltzmann constant and $\epsilon_{eff}$ is the effective emissivity of the SiN membrane.

The effective emissivity was calculated, following Kirchhoff's law, from the spectral absorptance. The absorptance of the membrane was estimated using the thin-film approximation:

\begin{equation}
    \alpha(\lambda)=1-\exp\left(-\frac{4\pi k(\lambda)h}{\lambda}\right)
\end{equation}

where $\lambda$  is the wavelength of the absorbed light, $k(\lambda)$ the extinction coefficient, and $h$  the membrane thickness. The wavelength-dependent absorptance was converted into an effective emissivity by weighting it with the black-body spectrum at the chuck temperature:

\begin{equation}
    \epsilon_{eff}=
\frac{\int \alpha(\lambda)B(\lambda,T_{chuck}),d\lambda}
{\int B(\lambda,T_{chuck}),d\lambda}
\end{equation}

where $B(\lambda,T)$ is Planck's black-body radiation distribution.

The temperature distribution calculated with the model was used to evaluate the steady-state temperature of the membrane during annealing.
 The results are reported in Fig.~\ref{fig:comsol_sim} for a SiN membrane simulated considering only thermal conduction (blue line) and conduction including radiative contributions (orange line). The latter significantly increases the temperature at the center of the structure. The main parameters used for the simulation are reported in Table~\ref{tab:simulation_parameters}.

\begin{table}[h]
\centering
\caption{Parameters used in the thermal simulation.}
\label{tab:simulation_parameters}

\begin{tabular}{lc}
\hline
Parameter & Value \\
\hline
$L$ & $1~\mathrm{mm}$ \\
$h$ & $30~\mathrm{nm}$ \\
$T_{chuck}$ & $1000\,^{\circ}\mathrm{C}$ \\
$\varepsilon_{eff}^{\dagger}$ & $4.55\times10^{-3}$ \\
$k_T$ & $3~\mathrm{W\,m^{-1}\,^{\circ}C^{-1}}$ \\
\hline
\end{tabular}
\\[2pt]
{\footnotesize
$^\dagger$ Calculated from the UV--Vis and infrared SiN spectra reported in Refs.~\cite{biliaev2022,kischkat2012}.
}

\end{table}

In plain membranes, dissipation occurs closer to the edges, where the temperature is higher and closer to the value of the frame and the chuck. Conversely, for the FTIR spectrum as well as tensile stress discussed in the main text, the average temperature is instead the relevant parameter, as it determines for example the overall reduction of surface OH groups density after annealing, although a spatial gradient of the density across the membrane is likely present.

\section{Thermal treatment reversibility with air exposure}

\begin{figure}
\centering
    \includegraphics[width=\columnwidth]{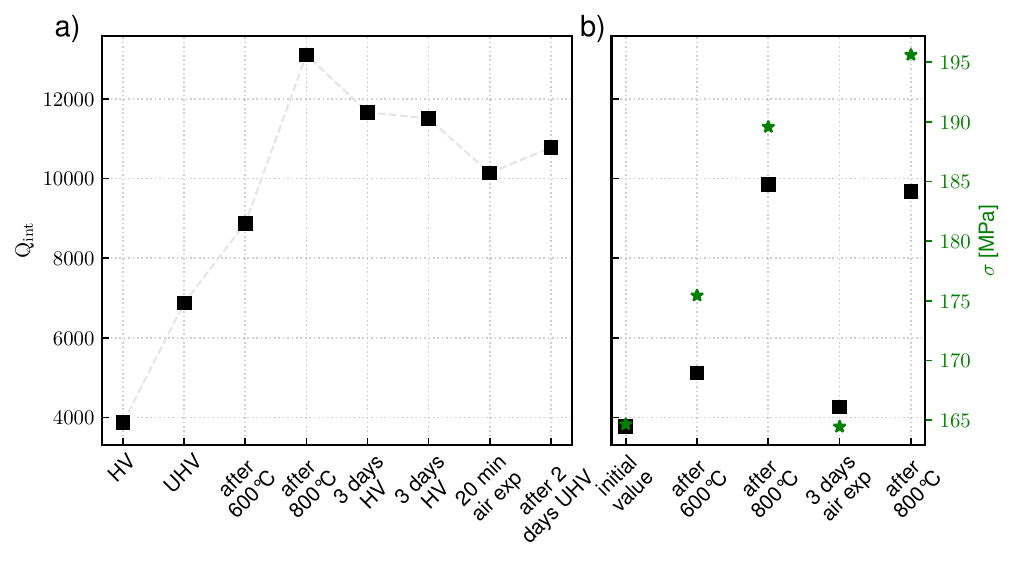}\\[1ex]
    \includegraphics[width=\columnwidth]{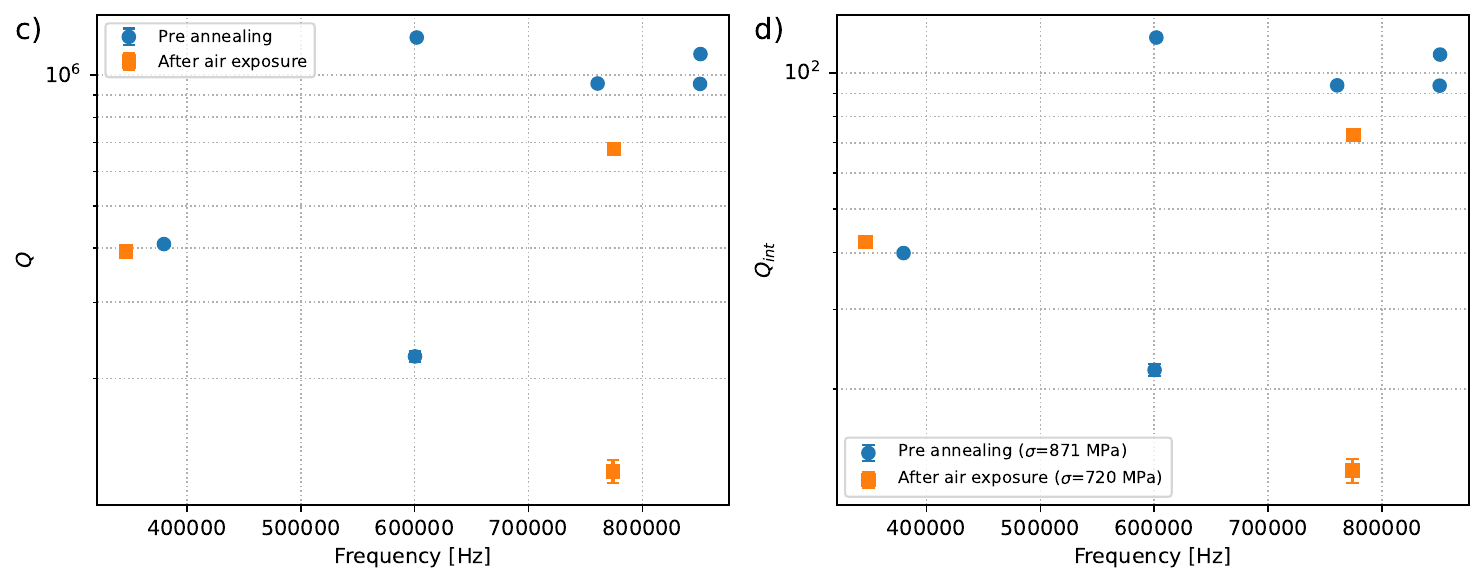}
    \caption{\textbf{(a)} $Q$ measurements of a 50LS sample after
    3 and 6 days in the load lock. After 20 minutes of air exposure, the
    sample is transferred back into the main chamber and measured
    immediately, and again after 2 days in UHV.
    \textbf{(b)} Different sample from the same wafer as in (a):
    $\sigma$ and $Q_\mathrm{int}$ after annealing at 600 and 800$^\circ$C
    for 1 hour, after 3 days of air exposure, and after a final annealing
    at 800$^\circ$C following the air exposure.
    \textbf{(c)} 10HS sample: quality factors of all measured modes in the
    untreated state and after a few months of air exposure following
    thermal treatments up to 1200$^\circ$C.
    \textbf{(d)} $Q_\mathrm{int}$ for the same sample as in (c).}
    \label{fig:reversibility_air_exp_suppl}
\end{figure}

\begin{figure}
\centering
    \includegraphics[width=\columnwidth] {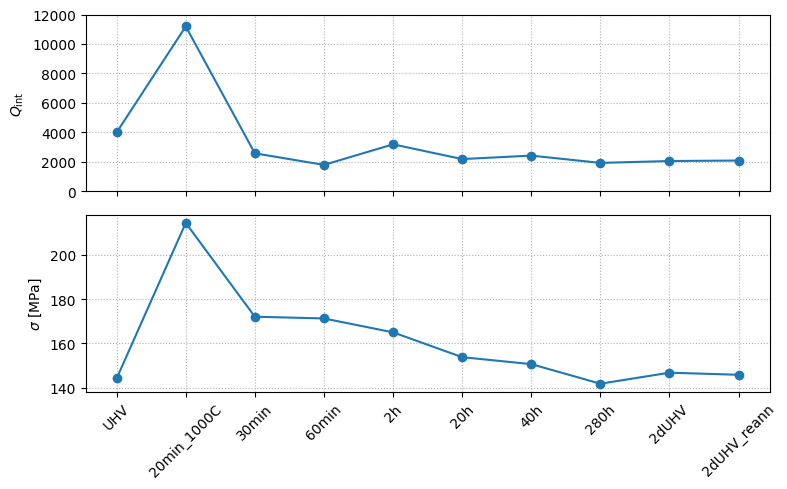}
\caption{complete history of a sample 30LS. After the thermal treatment of 20 minutes at 1000°C with a ramp up/down of 40°C/min, the sample is exposed to a humid environment (RH>95$\%$, 25$^{\circ}$C) and remeasured in UHV after different times. After 280h of exposure the sample was measured immediately and after 2 days in the chamber to see some differences and finally annealed at 150°C for 10 minutes to test if the water that caused the modification of the sample was just adsorbed or reacted chemically with the surface groups.}  
    \label{fig:supp_complete_historyUHV49}
\end{figure}

Before moving the thermally treated samples to the FTIR set-up, the effect of different exposure condition was tested on a 50LS sample (see Fig.~\ref{fig:reversibility_air_exp_suppl}). In Fig.~\ref{fig:reversibility_air_exp_suppl}a is reported the effect of moving the sample from the HV setup to the UHV , which leads to an increase of the quality factor by a factor of about $75\%$ reaching approximately 7000. After that, the first annealing at $600^{\circ}$C for 1 hour  further improves the $Q_{\mathrm{int}}$ to $9000$ and even more to $13000$ after another hour at $800^{\circ}$C. The Qs appear to be relatively stable after 3 and 6 days in HV, dropping just by $10\%$ slightly below 12000. The drop after just 20 minutes in air is in comparison much higher, reaching approximately $Q_{int}=10000$. The value increases again after 2 days in the UHV chamber, indicating that the process is partially reversible for this timescale of air exposure. 
The Fig.~\ref{fig:reversibility_air_exp_suppl}b instead presents the effect on $Q_{\mathrm{int}}$ and $\sigma$ of a 3 days air exposure on the annealed 50LS sample at $800^{\circ}$C for 1 hour. The sample is then re-annealed in the same conditions. The Fig.~\ref{fig:reversibility_air_exp_suppl}b strongly suggests a correlation between stress and intrinsic dissipation variation generated by thermal treatment. Fig.~\ref{fig:reversibility_air_exp_suppl}c,d show the reversibility of the
10HS sample discussed in the main text. Although only three modes could be
measured, their quality factors are comparable to the values measured before
thermal treatment, confirming the full reversal of the enhancement upon air
exposure. The stress instead ends up below its pre-treatment value, as the
result of two subsequent reductions: the treatment cycles up to
$1200^{\circ}$C irreversibly relax the tensile stress through structural
relaxation of the amorphous network, and the
subsequent air exposure removes the remaining condensation-induced tensile
surface-stress contribution through rehydroxylation.
A similar exposure experiment but with a humidity close to saturation at 25$^{\circ}$C was conducted on a  30LS sample, this time the sample was measured in UHV after 0.5, 1, 2, 20, 40, and 280 hours of exposure (Fig.~\ref{fig:supp_complete_historyUHV49}). Later, the sample was first kept stored  in the UHV chamber for 48 hours. This has not affected the $Q_{\mathrm{int}}$ but has slightly enhanced  the stress  ($\approx$5MPa). Finally, possible effects on mechanical properties of weakly adsorbed water were tested by a thermal treatment of 10 minutes at $150^{\circ}$C. The latter has not affected the discussed mechanical parameters, suggesting that in the UHV environment the dissipation of the  samples under investigation is not limited by adsorbed water. Instead, stress might be more sensitive \cite{stgelais_surfenergy2025}. 


\section{Modification induced by the storage and vacuum environment}
\begin{figure}
\centering
    \includegraphics[width=\columnwidth] {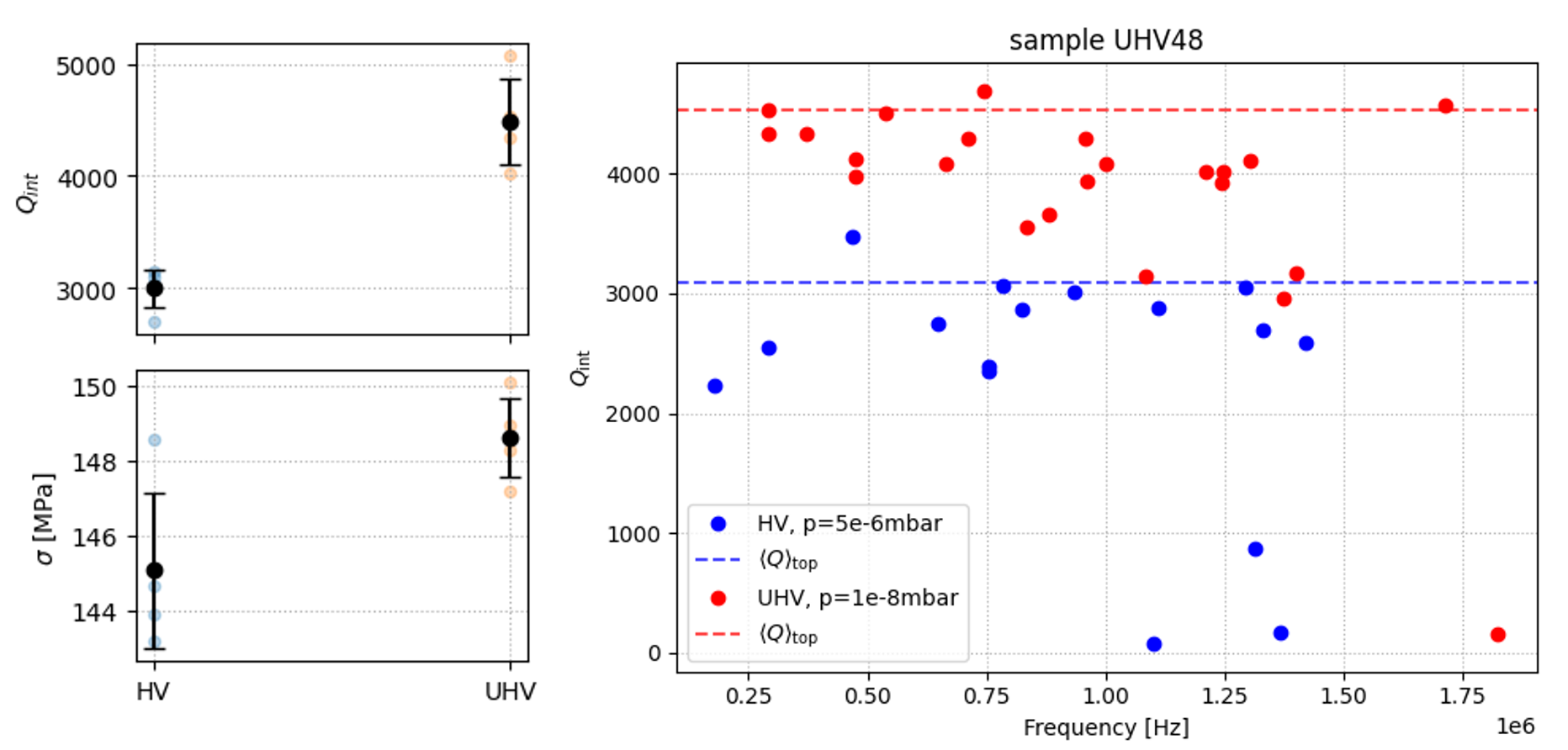}
    \caption{ left: $Q_{\mathrm{int}}$ and stress measured in high vacuum (HV, p=5e-6mbar) and ultra-high vacuum (UHV, p=1e-8mbar) for the same set of 4 samples 30nm-thick, 200MPa of prestress. right: complete dataset for one sample for HV (blue) and UHV (red). the dashed lines are calculated using the 5 highest value of $Q_{\mathrm{int}}$.  }
    \label{fig:supp_differenceHV_UHV}
\end{figure}

The Fig.~ \ref{fig:supp_differenceHV_UHV} presents on the left the $Q_{\mathrm{int}}$ and $\sigma$ measured in high vacuum (HV, p=5e-6mbar) and ultra-high vacuum (UHV, p=1e-8mbar) for the same set of 4 samples 30LS. The difference in $Q_{\mathrm{int}}$ cannot be fully explained by gas damping and might be due to a remaining sub-monolayer of water adsorbed on the surface from the ambient humidity. On the right a complete dataset of one sample in HV (blue) and UHV (red) is presented. The dashed lines are calculated using the 5 highest value of $Q_{\mathrm{int}}$. These data suggest that gas damping is not responsible for the difference measured in the two different vacuum conditions, because the $Q_{\mathrm{int}}$ measured in HV do not increase linearly after the first modes. The phenomenon is not yet fully understood but is one of the reasons why the UHV environment was chosen for carrying out the annealing experiments. The difference in stress could also be in principle explained by a difference in surface energy induced by adsorbed water \cite{stgelais_surfenergy2025}. Despite that, a contribution could also rise from the different temperature of the two vacuum chambers  and the different laser powers used for the readout.
\begin{figure}
\centering
    \includegraphics[width=\columnwidth] {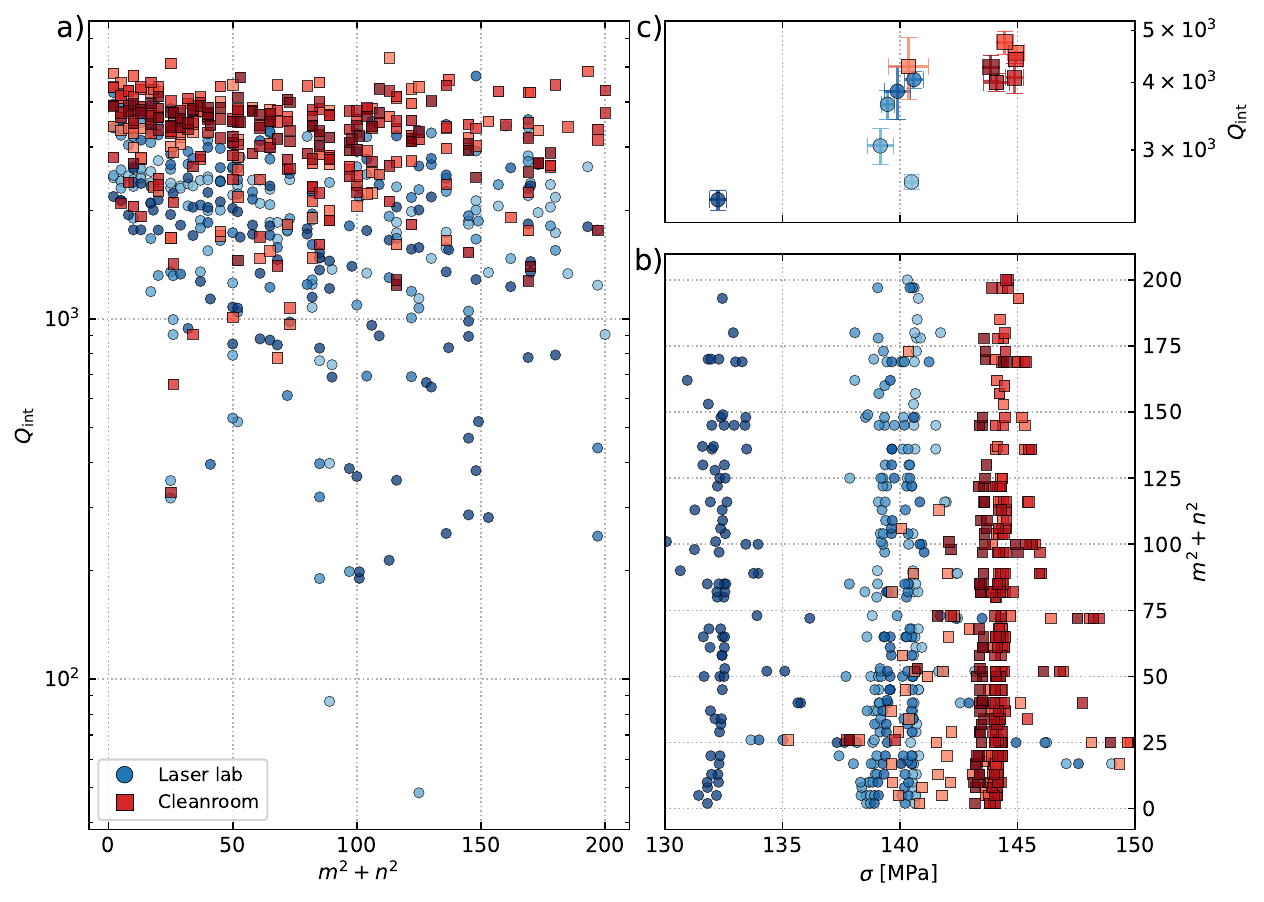}
\caption{Effect of different storage environment  approximately one year after. Samples were
stored either in the cleanroom (red squares) or in the laser lab
(blue circles). (a)~Quality factors of all measured modes as a
function of the mode number. (b)~Tensile stress extracted from each
resonance frequency.
(c)~Per-sample summary: the
two storage groups form distinct clusters, with cleanroom-stored samples
exhibiting both higher stress and higher
$Q_\mathrm{int}$.}  
    \label{fig:supp_storage}
\end{figure}

Fig.~\ref{fig:supp_storage} presents the mechanical properties of
as-received 30LS membranes approximately one year after fabrication,
grouped by storage environment: cleanroom (red squares) and laser lab (blue
circles). The complete set of measured quality factors is shown in
Fig.~\ref{fig:supp_storage}a, and the stress extracted from the resonance
frequencies via the dispersion relation (Eq.~3 of the main text) in
Fig.~\ref{fig:supp_storage}b. The clustering of the two groups is well captured in Fig.~\ref{fig:supp_storage}c, where each sample is represented
by the mean of its five highest $Q_\mathrm{int}$ values.

Cleanroom-stored samples show $\sigma = 143.9 \pm 0.6$~MPa and
$Q_\mathrm{int} = 4340 \pm 100$ (mean $\pm$ standard error, $N = 7$),
against $\sigma = 138.6 \pm 1.3$~MPa and $Q_\mathrm{int} = 3270 \pm 280$
($N = 6$) for the laser-lab set. Both differences are statistically
significant at the level of the group means: $5.3 \pm 1.4$~MPa in stress
(Welch's $t$-test, $p = 0.007$) and $1060 \pm 290$ in $Q_\mathrm{int}$
($p = 0.01$). Cleanroom storage is thus associated with both higher tensile
stress and higher intrinsic quality factor, suggesting that the samples age
differently depending on the storage environment. Whether the difference originates from the
distinct humidity levels or from airborne contamination in the two
environments remains open.
\section{XPS complete data}

\begin{figure}
    \centering

    \subfloat[]{
        \includegraphics[width=0.95\textwidth]{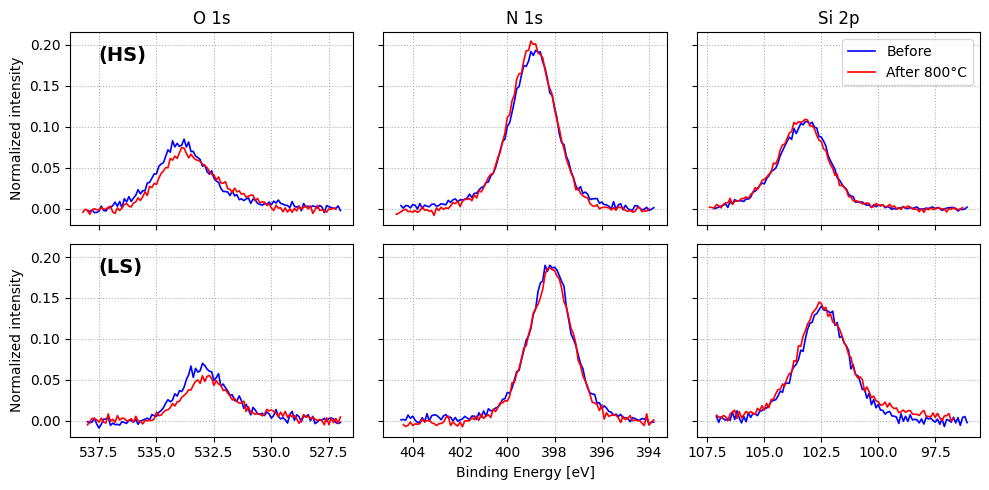}
        \label{fig:XPS1_suppl}
    }
    \vfill
    \subfloat[]{
        \includegraphics[width=0.95\textwidth]{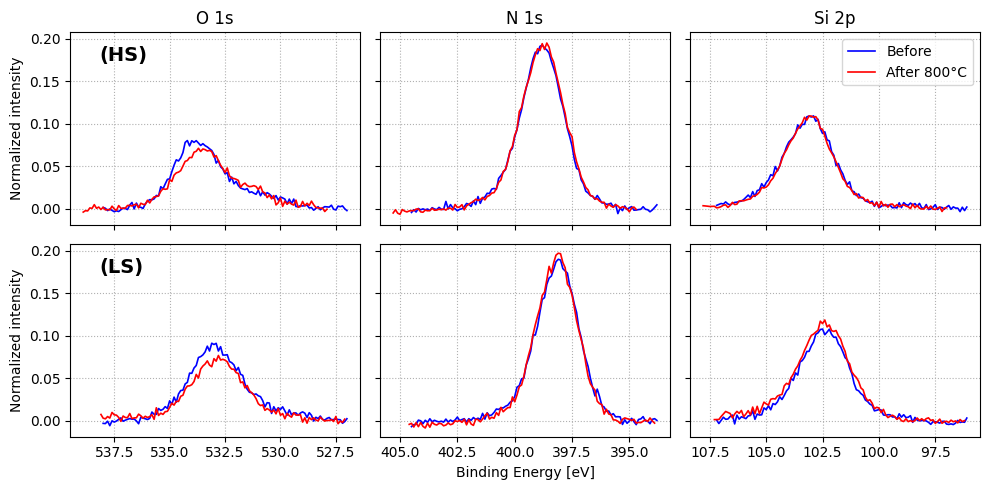}
        \label{fig:XPS2_suppl}
    }

    \caption{
    (a) Measurements of O 1s, N 1s, and Si 2p for  high (up) and low (down) stress 30nm-samples taken before (blue) and after (red) thermal treatment at 800$^{\circ}$C for 10 minutes, respectively. The data for oxygen are also presented in the main text. (b) same as (a) but for different samples. the data are shifted and normalized for the sum of the three areas, for each sample. 
    }
    \label{fig:XPS_supp}
\end{figure}

All XPS measurements were performed in a UHV chamber at a base pressure of $5\cdot10^{-11}$ mbar. Monochromatic Al K$\alpha$ radiation (E$_{\gamma}$ = 1486.65 eV) was supplied with an incidence angle of 45$^\circ$ by a SPECS XR50 source equipped with an aluminum anode. Photoelectrons were collected from the sample surface at a 90$^\circ$ take-off angle using a SPECS PHOIBOS 150 hemispherical analyzer and detected with a SPECS 2D-CCD detector. The samples were measured with XPS twice: once right after insertion into the UHV chamber and once after being annealed at 800 $^\circ$C for 10 min. The heat treatment was performed in situ by electron-beam heating of the sample holder, while temperature was manually controlled using an infrared pyrometer (DIAS DGE10N, $\pm$2$^\circ$C accuracy) and a temperature ramp close to 1 $^\circ$C/s.  
The spectra are normalized to the sum of the O~1s, N~1s, and Si~2p peak intensities. Additionally, the data are shifted by the offset in the N~1s peak positions to account for sample charging effects.
The complete set of XPS data, composed of two low stress and two high stress samples, measured before and after annealing at  800$^{\circ}$C for 10 minutes. The results are reported in Fig.~\ref{fig:XPS_supp}. The measured surface was the one not etched with KOH.
The atomic concentration was calculated using:
\begin{equation}
   at.\%_i =
    \frac{A_i / RSF_i}{\sum_j A_j / RSF_j} \cdot 100 
\end{equation}

where $A_i$ is the integrated peak area after background subtraction and $RSF_i$ is the relative sensitivity factor of element $i$. The calculated data for the four samples are reported in Tab.\ref{tab:compositional_suppl}. The only notably difference between before and after thermal treatment is a moderate reduction in the oxygen content, much less than what was found after wet etching of the oxide layer with HF \cite{giesriegl2026hf}.  Interestingly, the composition of low and high stress does not change considerably. Taking into account the data of $Q_{\mathrm{int}}$ and $\sigma$ reported in the main text, the XPS data suggest that the oxide layer thickness does not correlate with the change in dissipation and stress observed with thermal treatments.
\begin{table}
    \centering
    \caption{Calculated compositional percentage}
    \label{tab:compositional_suppl}
    \scalebox{1}{
        \begin{tabular}{|l|c|c|c|l|l|l|}\hline
 & \multicolumn{3}{|c|}{Before}& \multicolumn{3}{|c|}{After}\\\hline
            
             &O 1s& N 1s& Si 2p& O 1s& N 1s&Si 2p\\\hline
            
             HSa&11.6& 38.5& 50.0& 10.4& 39.2&50.3\\\hline
             HSb&12.7& 38.3& 49.0& 11.7& 37.8&50.5\\\hline
             LSa&8.5& 35.7& 55.8& 7.8& 34.9&57.4\\\hline
             LSb&14.0& 35.9& 50.1& 11.1& 32.6&56.3\\\hline

        \end{tabular}
    }
\end{table}

\section{FTIR extra data}

\begin{figure}
\centering
    \includegraphics[width=0.9\columnwidth] {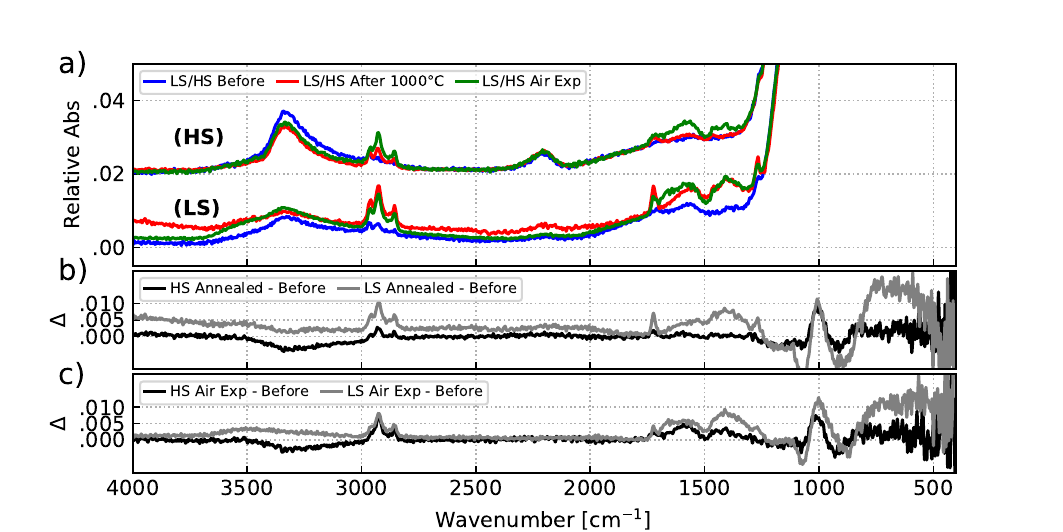}
    \caption{(a) Second set of data for HS and LS SiN-30nm-sample  before  (blue), after thermal treatment at $1000^{\circ}$C (red) and, after 1 week for HS and 4 weeks for LS of air exposure. (b) Difference After - Before thermal treatment for HS (black) and LS (grey). (c) Difference After air exposure - Before thermal treatment for HS (black) and LS (grey). }
    \label{fig:FTIR_suppl}
\end{figure}

In Fig.~\ref{fig:FTIR_suppl} are plotted the FTIR measurements of one other 30HS and 30LS sample. In Fig.~\ref{fig:FTIR_suppl}a are plotted the spectra before thermal treatment (blue), after thermal treatment at 1000$^{\circ}$C (red) and after one week of air exposure for the HS and after four weeks for the LS. In Fig.~\ref{fig:FTIR_suppl}b the different spectra between the treated and untreated is plotted for HS (black) and LS (gray). Fig.~\ref{fig:FTIR_suppl}c reports on the other hand the difference between after air exposure and before thermal treatment. The 30LS after annealing has a baseline that alters the results and after air exposure the OH/NH peak is even higher than the untreated sample. This could be due to different condition in storage of the sample and before the treatment and after (the humidity in the FTIR lab was not controlled). In addition the peak at 1010$cm^{-1}$ does not show a significant increase after air exposure, conversely of what observed for the sample discussed in the main text. The 30HS is just partially recovered, probably because the air exposure time was not enough.  

\section{Refractive index vs thickness}

\begin{figure}
\centering
    \includegraphics[width=0.9\columnwidth] {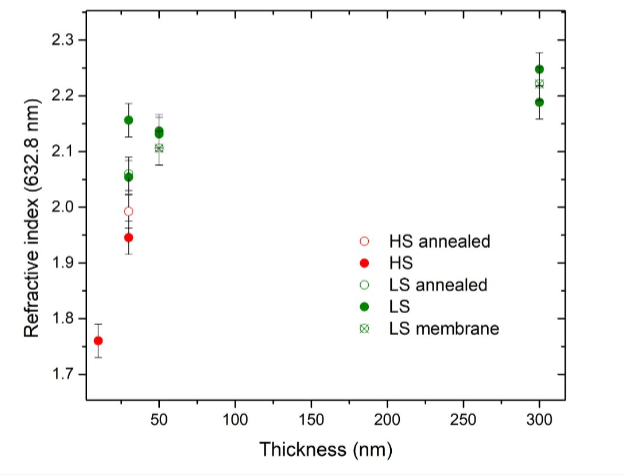}
    \caption{Refractive index comparison of SiN sample with 200MPa (green) and 1100MPa (red) Measured with ellipsometry at 632.8nm. Empty markers are from annealed samples at 1000$^{\circ}$C for 1 hour, but after several days of air exposure. The error bars are calculated from the fitted model.}
    \label{fig:ellipsometry_suppl}
\end{figure}

The refractive index ($n$) of 10, 30, 50, and 300 nm samples fabricated at 200 MPa and 1100 MPa and discussed in the text was measured by ellipsometry. The results, reported in Fig.~\ref{fig:ellipsometry_suppl}, indicate that $n$ is essentially independent of the membrane thickness down to 30 nm, within the experimental scatter. The 30 nm  HS sample exhibits a slightly lower refractive index than the corresponding LS samples, while the thermal treatment at 1000$^{\circ}$C for 1 hour has no significant effect on $n$. In contrast, the 10 nm HS sample shows a substantially lower refractive index.

This observation can be interpreted using an effective-medium model consisting of a SiN core covered by oxide layers on both surfaces. Let $n_{SiN}=2$ and $n_{ox}=1.4$ be the refractive indices of SiN and the oxide, respectively, and let $n_{eff}$ be the effective refractive index measured by ellipsometry. The oxide thickness can then be estimated as

\begin{equation}
    t_{ox}=\frac{h}{2}\frac{n_{SiN}-n_{eff}}{n_{SiN}-n_{ox}}
\end{equation}

where $h$ is the total membrane thickness. Using the measured refractive index of the 30 nm HS sample yields an oxide thickness of approximately $t_{ox}\approx1.25nm$, in good agreement with values reported in the literature \cite{antonius_paper, giesriegl2026hf}.

Assuming that this oxide thickness remains approximately unchanged for the 10 nm sample and that the porosity is uniformly distributed throughout the sample, the porosity fraction $x$ can be estimated as

\begin{equation}
    x=1-\frac{h(n_{eff}-1)}{(h-2t_{ox})n_{SiN}+2t_{ox}n_{ox}-h}\approx0.12 
\end{equation}

The estimated porosity of approximately 12$\%$, likely originating during fabrication, could account for the low quality factors measured for the 10 nm membranes compared with the expected linear scaling \cite{Silvan2014_evidence_of_surf_loss,benga2026determination}. A reduced material density and increased internal surface area associated with porosity are expected to enhance mechanical dissipation, providing a plausible explanation for the observed deviation.

\bibliographystyle{elsarticle-num}
\bibliography{biblio}